\documentclass[a4paper,11pt]{article}
\pdfoutput=1

\usepackage{jcappub}
\usepackage{amssymb}
\usepackage{lipsum}
\usepackage{amsmath}
\usepackage{graphicx} 
\usepackage{xcolor}

\usepackage[T1]{fontenc}

\newcommand{\Trh}{T_{\rm RH}}
\newcommand{\noi}{\noindent}
\newcommand{\beq}{\begin{equation}}
\newcommand{\eeq}{\end{equation}}
\newcommand{\bea}{\begin{eqnarray}}
\newcommand{\eea}{\end{eqnarray}}

\title{\boldmath Searching for UFOs from the early universe: direct detection prospects for relativistically decoupling dark matter}

\author[a,1]{Stephen E. Henrich,\note{Corresponding author.}}
\author[b]{Yann Mambrini,}
\author[a]{and Keith A. Olive}

\affiliation[a]{William I. Fine Theoretical Physics Institute, School of Physics and Astronomy, University of Minnesota,\\
Minneapolis, Minnesota 55455, USA}

\affiliation[b]{Universit\'e Paris-Saclay, CNRS/IN2P3, IJCLab,\\
91405 Orsay, France}

\emailAdd{henri455@umn.edu}
\emailAdd{mambrini@ijclab.in2p3.fr}
\emailAdd{olive@umn.edu}

\abstract{Particles that decouple relativistically from the Standard Model bath during reheating represent a versatile class of well-motivated cold dark matter 
candidates. In fact, 
ultrarelativistic decoupling ($T_{\rm FO}\gg m_\chi$) 
is quite generic for beyond the Standard Model (BSM) heavy portal 
interactions with strong couplings and relatively low reheating temperatures. In this work, we study the direct detection prospects for ultrarelativistically frozen-out (UFO) candidates, using $Z'$-portal dark matter as a case study. Although typical UFO cross sections are suppressed by a heavy mediator mass scale, we find that experiments such as LZ, XENONnT, PandaX, and DarkSide-50 have already excluded a large portion of the UFO parameter space and there remains viable space 
above the neutrino fog for $0.4 \text{ GeV} \lesssim m_{\rm DM}\lesssim 1$~TeV.  
Moreover, 
SuperCDMS SNOLAB, which is expected to begin collecting data in 2026, should 
access a large region of UFO parameter space in the 0.5-10 GeV mass range. 
For heavy BSM portal 
interactions ($M\gtrsim 1$~TeV), UFOs are typically more accessible to detection than freeze-in candidates due to the comparatively larger cross sections. We also carefully delineate 
regions of parameter space with degeneracy between UFO and non-relativistic freeze-out. In sum, UFOs are 
attractive candidates for ongoing and next-generation dark matter detection 
experiments in a looming post-WIMP era.}

\begin{document}

\begin{flushright}
{UMN--TH--4527/26, FTPI--MINN--26/09} \\
May 2026
\end{flushright}

\maketitle
\flushbottom

\section{Introduction}

Since Fritz Zwicky's seminal work \cite{Zwicky:1933gu}, the nature of dark matter has remained 
a mystery. Various candidates for particle dark matter began to emerge in the second half of the 20th century. While some were motivated 
by constructions such as supersymmetry \cite{Go1983,ehnos}, others were driven by a minimalist principle \cite{SS,Gunn:1978gr,Silveira:1985rk,McDonald:1993ex,Burgess:2000yq,Davoudiasl:2004be,Cirelli:2005uq}. 
Among the first simple models of dark matter was a light neutrino with an eV-scale mass \cite{SS}. 
This was the original ultrarelativistic freeze-out (UFO) candidate, as a neutrino with Standard Model (SM) electroweak interactions will freeze-out at a temperature of roughly 1 MeV. However, the SM neutrino is an example of hot dark matter \cite{Bond:1983hb} and due to free streaming \cite{free}, produces too much large scale structure \cite{nu3}. Indeed the latter constraints are so strong that the sum of light neutrino masses is limited to $\Sigma m_\nu \lesssim 0.1$~eV \cite{Planck:2018vyg,Brieden:2022lsd} whereas as dark matter candidates we must have $\Sigma m_\nu \simeq 11$~eV \cite{Gershtein:1966gg,Cowsik:1972gh,Szalay:1974jta}. UFOs as dark matter candidates were therefore promptly discarded.

On the other hand, a 4th generation SM-like neutrino (or heavy neutral lepton) was initially an ideal dark matter candidate \cite{Gunn:1978gr}. Its relic density is computed from non-relativistic freeze-out \cite{hut} and
an acceptable abundance of this cold dark matter candidate is found for $m_\nu \sim 4$~GeV. However, a 4th generation neutrino with mass less than $M_Z/2$
has been precluded by measurements of the width of the $Z$ gauge boson \cite{DELPHI:2003dlq}  and $e^+e^-$ scattering \cite{Janot:2019oyi}. 

There are of course many other examples of weakly interacting massive particles (WIMPs) whose relic density is determined by
 the "WIMP miracle" paradigm. This includes the plethora of supersymmetric candidates \cite{ehnos}.   Because of their electroweak scale couplings, it seemed that rapid discovery was ensured. 
An electroweak interaction strength ensures that the dark matter $\chi$ maintains equilibrium with the primordial 
plasma during the radiation-dominated era for temperatures, $T > m_\chi$.  At lower temperatures,  dark matter is depleted via annihilations until
freeze-out, which occurs when the annihilation rate, $\Gamma_A \propto n_\chi \langle \sigma v \rangle$ falls below the expansion rate, $H \propto T^2/M_P$, where $M_P$ is the Planck mass.  For electroweak strength interactions, non-relativistic freeze-out typically occurs when $m_\chi/T \sim 25$ and the yield $Y_\chi^0=\frac{n^0_\chi}{n_\gamma^0} \simeq (m_\chi/T)^{3/2} e^{-m_\chi/T} \lesssim 10^{-9}$, where $n^0_\gamma$ is the present photon density.  For $m_\chi \sim 1$ GeV the relic density is roughly $\Omega_\chi h^2 = \rho_\chi h^2/\rho_c \sim 0.1$ as is needed for standard $\Lambda$CDM cosmology.  
However, the  
absence of detection in experiments such as XENONnT \cite{XENON:2025vwd}, PandaX \cite{PandaX:2024qfu}, or 
LUX-ZEPLIN \cite{LZ:2024zvo,LZ:2025igz} has placed many WIMP candidates under pressure.

An alternative to the WIMP
is a particle such as the gravitino with Planck-suppressed interactions which was never in thermal equilibrium with the radiation bath \cite{ehnos,grav}. 
In this case, dark matter production occurs continuously from the scattering of SM particles in the thermal bath, but never so efficiently as to produce an equilibrium abundance of dark matter. Typically, the dark matter yield is $Y \sim \Gamma_P/H \sim \Trh/M_P$, where $\Gamma_P$ is the dark matter production rate, and $\Trh$ is the reheating temperature after inflation. 
This type of candidate has been generalized to a FIMP for Freeze-In Massive Particle, or Feebly Interacting Massive Particle \cite{fimp,Bernal:2017kxu}. 

The suppression of the FIMP production rate may be due to feeble couplings, or typical gauge-strength couplings but whose interactions are mediated by intermediate scale gauge or Higgs bosons as in the NETDM scenario \cite{Mambrini:2013iaa,Mambrini:2015vna}. Another possibility is ``freeze-in at stronger coupling" \cite{Cosme:2023xpa,Arcadi:2024obp},
where the feeble coupling is replaced
by a Boltzmann suppression of the particles in the bath, $e^{-\frac{m_\chi}{T}}$, 
if the mass of the DM lies above the reheating temperature.
We must also distinguish between two types of FIMP, depending on the temperature dependence of the effective cross section, which we can parameterize as $\sigma \propto T^n$ in the relativistic regime ($T\gg m_\chi$). In a radiation dominated universe with $H \propto T^2/M_P$ and $n>-1$, production occurs in the early stages, when the temperature of the thermal bath is at its highest. These are UV (UltraViolet) FIMPs \cite{Elahi:2014fsa,Bernal:2019mhf,Bernal:2025fcl}. For the other possibility ($n<-1$) the dominant contribution to the final relic density occurs at lower temperature when $T \simeq m_\chi$. These are IR (Infrared) FIMPs . In an effective field theory, there is a clean mapping from the operator dimension to $n$. In particular, for an effective Lagrangian with a dimension $d$ operator $\mathcal{O}_d$ given by $\mathcal{L}_{\rm eff} \supset \frac{\mathcal{O}_d}{\Lambda^{d-4}}$, the relevant $2\rightarrow 2$ thermally averaged scattering cross sections in the relativistic regime scale as $\langle \sigma v \rangle \propto \frac{T^{2d-10}}{\Lambda^{2d-8}}$. This implies $n=2d-10$. For instance, dimension 6 operators from integrating out a heavy mediator correspond to $n=2$. As a result, IR FIMPs typically correspond to interactions proceeding via renormalizable operators while UV FIMPs correspond to non-renormalizable operators. This distinction will prove important later on.

For a successful FIMP, the feebleness of the effective coupling ensures 
an extremely low comoving number density for dark matter relative to that of SM radiation.
The paradox is that, while there is a growing tension surrounding the WIMP paradigm which stems 
from the absence of experimental detection, the difficulty in the discovery of a FIMP stems precisely from 
the tiny coupling required to avoid an overproduction of dark matter that would 
overclose the Universe.

Another important feature of the FIMP dark matter paradigm is the assumption of zero density as an initial condition. 
This is assumed without rigorous justification. 
Indeed, 
one can argue instead that this assumption is not justified as it 
is impossible to omit, for example, gravitational production. The amount of dark matter produced may depend on the details of reheating after inflation, and the reheating temperature, such that taking the initial abundance to be zero at $T=T_{\rm RH}$ is unjustified. Moreover, gravitational inflaton scattering and/or thermal scattering of inflaton decay products prior to reheating will contribute to the initial density of any dark matter particle \cite{Mambrini:2021zpp,Clery:2021bwz}. For scalar dark matter, there are in addition long wave-length modes which contribute the final dark matter density \cite{Cosme:2020nac,Garcia:2025rut}.
Thus in the absence of a phase of thermal equilibrium, one must be careful about the specifications of initial conditions. This problem of initial conditions obviously does not exist for WIMPs, since 
thermal equilibrium erases the candidate's entire past production history. 

Various options 
have recently been proposed to address the problems associated with the freeze-in and freeze-out mechanisms, many of which require a more detailed study of reheating \cite{Henrich:2024rux}. As noted above, with freeze-in at strong coupling FIMP production is minimized due to Boltzmann suppression, while maintaining a coupling to Standard Model particles strong enough to be observable.

Another option, discussed in \cite{Henrich:2025sli,Henrich:2025gsd}, looks closely at UFO during the reheating phase, rather than during the 
radiation-dominated era. This involves applying the known neutrino freeze-out mechanism, but during reheating when the thermal bath is still being produced by 
inflaton decays. The problem of initial 
conditions is avoided by the thermalization of the dark matter candidate. As it freezes-out while still relativistic, there is no Boltzmann suppression of its density.  However, after freeze-out, the dark matter density is diluted as the thermal plasma continues to be created, thus ensuring a sufficiently low yield $Y_\chi^0$. Moreover, its free streaming length is also modified due to the dynamics of expansion during reheating, making it a candidate for cold dark 
matter, in contrast to the neutrino which if massive acts as hot dark matter. While we consider relativistic freeze-out from the Standard Model during reheating in this work, relativistic freeze-out can also be rendered viable (i.e. by evading warmness constraints) if the dark matter decouples from a secluded sector \cite{Hambye:2019tjt,Hambye:2020lvy,Coy:2021ann,Coy:2024itg,ArcadiLebedev:2019,SterileNeutrinoFOLebedev} which we do not consider here.  An interesting 
analysis comparing WIMPy, FIMPy and UFOy dark matter has recently been carried out in \cite{Chakraborty:2026chp}.

As in the case of a FIMP, the production of UFO DM can either occur in the UV or IR, as discussed in detail in \cite{Henrich:2025gsd}.  This distinction depends mainly on two factors: the efficiency 
of reheating (how the temperature $T$ evolves as a function of the scale 
factor $a$) and the efficiency of dark matter production from the 
thermal bath 
(how the particle production rate {\it following} the relativistic freeze-out evolves as a function of temperature, which is given by the parameter $n = 2d-10$, where $d$ is the dimension of the operator responsible for the DM production).
In our analysis here, reheating is assumed to arise from the decay of 
the inflaton $\phi$, whose potential, following the inflationary period, is 
approximated by a quadratic potential, $V(\phi)\propto \phi^2$.
In this case, UV production only occurs for $n>6$. In the models considered here where the interactions are mediated by a heavy vector boson, $n=2$, and out-of-equilibrium production will continue after freeze-out until the temperature drops to a characteristic IR scale, namely $m_\chi$ or $\Trh$, whichever is greater\footnote{Note that while $T_{\rm RH}$ is regarded as a UV scale when dark matter is studied during radiation domination, it may be regarded as an IR scale when dark matter production occurs during reheating, since $T_{\rm RH}$ will often be the lower limit of the temperature integration. The maximum temperature achieved during reheating, $T_{\rm max}\gg T_{\rm RH}$, is instead often the relevant UV scale.}.
For extensions to this scenario, we refer the reader to \cite{Henrich:2025gsd}.

Because UFO dark matter comes into thermal equilibrium it also has the advantage of potentially interacting  sufficiently strongly with Standard Model particles to be observable in direct detection experiments, allowing one to be able to place constraints on the model. This is the goal of our present study. The paper is organized as follows.  In Section \ref{approach}, we provide the basic inputs to our calculations. This includes the Lagrangian for the dark matter and SM couplings to the massive vector, $Z'$. We also provide the annihilation and scattering cross sections for both a (Dirac) fermion and complex scalar DM candidate, both labeled as $\chi$. 
In Section \ref{relic}, we compute the relic density in the $(m_\chi,\Trh)$ plane for various values of $M_{Z'}$ and delineate the FIMP/UFO/WIMP regimes. Our main results are given in Section \ref{results}, where we illustrate the regions of the $m_\chi, \Trh, M_{Z'}$ parameter space which are accessible to current and future direct detection experiments. A summary of our results is provided in Section \ref{summary}.

\section{Approach}
\label{approach}

As in \cite{Henrich:2025pca}, we will focus on UFO dark matter which interacts with the Standard Model via a $U(1)'$ gauge boson, $Z'$, with mass $M_{Z'}\geq 1$~TeV. Our study can easily be 
generalized to scalar mediators. 
In this section, we review the effective interaction cross sections between dark matter and a nucleon through the exchange of a vector boson. We also supply the relevant annihilation cross sections. We will consider both fermion and scalar dark matter candidates.

It is worth noting that for the models we consider, dark matter production channels which are sourced by on-shell mediator states such as $Z'\rightarrow \chi \chi$ and $Z' Z' \leftrightarrow \chi \chi$ do not contribute significantly for the majority of our parameter space, despite the fact that $Z'$ will thermalize at high temperatures. This is due to the fact that $\chi$ will typically remain in equilibrium with the SM bath through the annihilation channel $\bar{f}f \leftrightarrow \chi \chi$ long after the $Z'\rightarrow \chi \chi$ and $Z' Z' \leftrightarrow \chi \chi$ channels become inactive, rendering their prior contributions negligible. This is largely because  we consider strong couplings (e.g. $V_\chi=V_f=1$) which ensures highly efficient $\bar{f}f \leftrightarrow \chi \chi$ annihilations (even after the $Z'$ equilibrium number density becomes Boltzmann suppressed) as well as ensuring that the $Z'$ particles closely track their equilibrium number density. In addition, the UFO parameter space always satisfies the hierarchy $M_{Z'}\gg m_\chi,m_f$, such that the equilibrium density of $Z'$ particle will be highly suppressed when the $\bar{f}f \leftrightarrow \chi \chi$ channel's production is at its peak. When couplings between the mediator and DM are much weaker, as in \cite{Becker:2023tvd}, freeze-in may proceed via decays of the mediator, while the $2\rightarrow2$ annihilation channel via an off-shell mediator in such a case is suppressed. Similarly, in the case where the SM-mediator coupling, $g_{\rm SM}$, is relatively small (as in \cite{Belanger:2024yoj}), the $2\rightarrow2$ annihilations are penalized relative to the mediator-to-DM decay channel, with the former proportional to $g_{\rm SM}^2g_{\chi}^2$ and the latter going as $g_{\chi}^2$. Here, we consider a distinct scenario where the $2\rightarrow 2$ annihilations via off-shell mediators will nearly always be dominant relative to any on-shell channels due to the strong couplings which ensure equilibration of $\chi$ is maintained after the mediator's equilibrium abundance becomes Boltzmann suppressed. For completeness, we include the $Z'\rightarrow \chi \chi$ and $Z' Z' \leftrightarrow \chi \chi$ channels in our analysis, as we did in \cite{Henrich:2025pca}; however, in the sections to follow we focus on the $\bar{f}f \leftrightarrow \chi \chi$ channel which dictates nearly all of the relevant UFO dynamics.

\subsection{Fermionic Dark Matter}

We start with a generic Lagrangian for a $Z'$ mediator with vector and axial vector couplings to Dirac fermion dark matter, $\chi$, and Standard Model fermions, $f$.
\begin{equation}
\mathcal{L} \supset \bar{\chi}(V_{\chi}\gamma^{\mu}+A_{\chi}\gamma^{\mu}\gamma_5)\chi Z^{'}_{\mu}+\sum_f \bar{f}(V_{f}\gamma^{\mu}+A_{f}\gamma^{\mu}\gamma_5)f Z^{'}_{\mu}\,.
\end{equation}

\subsubsection{Annihilation cross sections}

In our analysis below, we consider a pure vector coupling for spin-independent DM-nucleon scattering and a pure axial vector coupling for the case of spin-dependent scattering, which are frequently used as the standard benchmarks. Here we provide the relevant annihilation cross sections for each $2\rightarrow2$ process for completeness. The spin-averaged squared matrix element for the process $\overline{\chi}\chi\rightarrow \overline{f}f$ (assuming $f$ is Dirac) was given in \cite{Zheng:2010js,Henrich:2025pca}.  After integration over phase space, the annihilation cross section for the process $\overline{\chi}\chi\rightarrow \overline{f}f$ for pure vector coupling ($A_f=A_\chi=0$) is 
\begin{equation}
\sigma_{\overline{\chi}\chi\rightarrow \overline{f}f}(s) v_{\rm rel} = N_c^f V_f^2V_\chi^2\frac{\sqrt{1-\frac{4m_f^2}{s}}}{6\pi s}\frac{(s+2m_\chi^2)(s+2m_f^2)}{(s-M_{Z'}^2)^2+\Gamma_{Z'}^2M_{Z'}^2}, \label{fermionDMannCS}
\end{equation}
while the annihilation cross section for the pure axial vector coupling case ($V_f=V_\chi=0$) is
\begin{equation}
\sigma_{\overline{\chi}\chi\rightarrow \overline{f}f}(s) v_{\rm rel} = N_c^f A_f^2A_\chi^2 \frac{\sqrt{1-\frac{4m_f^2}{s}}}{6\pi s} \frac{s^2-4s(m_\chi^2+m_f^2)+28m_\chi^2m_f^2}{(s-M_{Z'}^2)^2+\Gamma_{Z'}^2M_{Z'}^2},
\end{equation} 
where $N_c^f$ is the color factor for the SM fermions $f$.
In the non-relativistic limit and with $s \ll M_{Z'}^2$, the s-wave parts of the cross sections are
\beq
\sigma v_{\rm rel} = \sum_f N_c^f V_f^2V_\chi^2 \sqrt{1-\frac{m_f^2}{m_\chi^2}} \frac{\left(2 m_\chi^2 + m_f^2\right)}{2\pi M_{Z'}^4}  
\, ,
\label{sannv}
\eeq
and
\beq
\sigma v_{\rm rel} = \sum_f N_c^f A_f^2 A_\chi^2 \sqrt{1-\frac{m_f^2}{m_\chi^2}} \frac{ m_f^2 }{2\pi M_{Z'}^4}\, ,
\label{sanna}
\eeq
respectively and the sum is over all 45 left and right handed SM fermions \footnote{This cross section assumes a Dirac fermion in the final state}. In principle, there could also be contributions from 
terms proportional to $V_\chi^2 A_f^2$, but we do not consider those here\footnote{The s-wave contribution from terms proportional to $A_\chi^2 V_f^2$ vanish, and the cross section term proportional to $V_\chi V_f A_\chi A_f$ vanishes at all partial waves.}.

\subsubsection{Spin-independent scattering cross section}

For a heavy $Z'$ with mass much greater than the momentum transfer
$k$, 
$(M_{Z'}^2\gg k^2)$, we can integrate out the heavy $Z'$ to obtain the effective Lagrangian with a four-fermion interactions
\begin{equation}
\mathcal{L}_{\rm eff}\!\supset\! \frac{1}{M_{Z'}^2} \sum_f \left(V_\chi V_f (\bar{\chi} \gamma^{\mu} \chi )(\bar{f} \gamma_{\mu} f) +A_\chi A_f (\bar{\chi} \gamma^{\mu}\gamma_5 \chi )(\bar{f} \gamma_{\mu}\gamma_5 f)+ ...\right)\,. 
\label{effLf}
\end{equation}

The cross section for the scattering of the dark matter $\chi$ on nucleons can be obtained from Eq.~(\ref{effLf}) by rewriting the Lagrangian
in terms of the nucleon operator, $N$, with a redefined coupling, $V_N$, 
\begin{equation}
\mathcal{L}_{\rm eff}\supset \frac{1}{M_{Z'}^2} \left(V_\chi V_N (\bar{\chi} \gamma^{\mu} \chi )(\bar{N} \gamma_{\mu} N)+ ...\right) \, ,
\end{equation}
where
\begin{equation}V_{N}=\sum_{q} V_q \,\langle N|\bar q\gamma^0 q|N\rangle\approx \sum_{q} V_q \,n_q^{(N)},\end{equation}
with $n_q^{(N)}$ being the total number of $q$-type quarks in the nucleon. For the vector current, the nucleon matrix element only counts valence quarks.

For DM-nucleon scattering experiments, we require the elastic DM-nucleon scattering cross section.
The amplitude for the vector interaction process is simply
\begin{equation}i\mathcal{M}=\frac{V_\chi V_N}{M_{Z'}^2}[\bar{u}_\chi(p_3)\gamma^{\mu}u_\chi (p_1)\cdot\bar{u}_N(p_4)\gamma_{\mu}u_N (p_2)]\,,\end{equation}
where $p_1, p_2$ ($p_3,p_4$) are in the incoming (outgoing) momenta. In the non-relativistic limit, the amplitude is dominated by the $\gamma^0$ term, such that we have 
\begin{equation}i\mathcal{M}=\frac{V_\chi V_N}{M_{Z'}^2}(2m_\chi)(2m_N)\,.\end{equation}
 Averaging initial spins and summing over final spins yields
\begin{equation}
\overline{|\mathcal{M}|}^{2}=16 V_\chi^2V_N^2\frac{m_\chi^2m_N^2}{M_{Z'}^4}.
\end{equation}
The $2\rightarrow2$ scattering cross section in the center of momentum frame can be written as 
\begin{equation}\sigma=\frac{S}{16 \pi s} \overline{|\mathcal{M}|^2} \, ,
\end{equation}
where $S$ is a symmetry factor.  
In the non-relativistic limit, $s\approx (m_\chi+m_N)^2$ and taking $S=1$, gives us 
\begin{equation}
\sigma^{\rm SI}_{\chi N}=\frac{V_\chi^2V_N^2 m_\chi^2 m_N^2}{\pi M_{Z'}^4(m_\chi+m_N)^2}=\frac{V_\chi^2V_N^2 \mu_{\chi N}^2}{\pi M_{Z'}^4}\,,\label{sigmaSI}
\end{equation}
where we have used the reduced mass $\mu_{\chi N}=\frac{m_\chi m_N}{m_\chi+m_N}$. For DM-proton and DM-neutron scattering respectively, we have 
\begin{equation}
V_p=\sum_{q} V_q \,\langle p|\bar q\gamma^0 q|p\rangle\approx \sum_{q} V_q \,n_q^{(p)}=(2V_u+V_d)\,,
\end{equation}
\begin{equation}
V_n=\sum_{q} V_q \,\langle n|\bar q\gamma^0 q|n\rangle\approx \sum_{q} V_q \,n_q^{(n)}=(V_u+2V_d)\,,
\end{equation} 
such that for universal couplings ($V_u = V_d\equiv V_q$) we have
$V_n = V_p= 3V_q$. 
We then obtain
\begin{equation}
\sigma^{\rm SI}_{\chi N}=\frac{9 V_\chi^2 V_q^2 \mu_{\chi N}^2}{\pi M_Z'^4}\,,
\label{Eq:sigmasi}
\end{equation}
when the momentum transfer $k$ goes to 0.

\subsubsection{Spin-dependent scattering cross section}

The axial vector part of the Lagrangian (\ref{effLf}) for the scattering on nucleons can be written as
\begin{equation}\mathcal{L}_{\rm eff}\supset \frac{1}{M_{Z'}^2} \left(A_\chi A_N (\bar{\chi} \gamma^{\mu}\gamma_5 \chi )(\bar{N} \gamma_{\mu}\gamma_5 N)+ ...\right) \, ,
\end{equation}
where
\begin{equation} 
A_N=\sum_q A_q \Delta q^{(N)} \, .
\end{equation}
with  $\Delta q^{(N)}$ the nucleon axial charges

The analogous spin-dependent DM-nucleon scattering cross section is 
\[\sigma^{\rm SD}_{\chi N}=\frac{3\mu_{\chi N}^2A_\chi^2 A_N^2}{\pi M_{Z'}^4}.\]
The axial charges are taken as follows \cite{Sato:2016tuz}:
\begin{align}
\Delta u^{(p)}&=+0.83 \pm 0.01, \hspace{0.5cm} \Delta u^{(n)}=-0.44 \pm 0.01  \\
\Delta d^{(p)}&=-0.44 \pm 0.01, \hspace{0.5cm} \Delta d^{(n)}=+0.83 \pm 0.01 \\
\Delta s^{(p)}&=-0.10 \pm 0.01, \hspace{0.5cm} \Delta s^{(n)}=-0.10 \pm 0.01
\end{align} 
Note that these values are taken from a recent analysis of spin dependent parton distributions from deep inelastic scattering (DIS). These indicate a relatively high contribution for $\Delta s$. Another study \cite{Ethier:2017zbq} which includes semi-inclusive DIS but drops SU(3) symmetry relations indicates a lower value though with a higher uncertainty, $\Delta s = -0.03 \pm 0.1$.
Depending on the value of $\Delta s$, $\Delta u$ and $\Delta d$ can be obtained from the combination $a_3 = \Delta u^{(p)} - \Delta d^{(p)} = 1.2753 \pm 0.0013$ which is very precisely measured from axial and vector contributions to neutron decay \cite{ParticleDataGroup:2024cfk} and $a_8 = \Delta u^{(p)} + \Delta d^{(p)} -2\Delta s = 0.585 \pm 0.025$ from the inclusion of hyperon $\beta$-decays \cite{AsymmetryAnalysis:1999gsr}. 
For the specific choice of universal axial vector quark couplings $A_q$, we find 
\begin{equation}\sigma^{\rm SD}_{\chi N}\simeq\frac{3\mu_{\chi N}^2A_\chi^2 A_q^2(0.29)^2}{\pi M_{Z'}^4} \, .
\label{sdsigma}
\end{equation}

\subsection{Scalar Dark Matter}

As an alternative to fermionic DM, we can also consider a complex scalar DM candidate, also denoted by $\chi$, charged under the $U(1)'$ gauge group. In this case the relevant terms in the Lagrangian are
\begin{equation}
\mathcal{L} \supset i V_\chi Z'^{\mu}\chi^{\dagger} \overleftrightarrow{\partial_\mu} \chi  -m_\chi^2 \chi^\dagger \chi +\sum_f \bar{f}(V_{f}\gamma^{\mu}+A_{f}\gamma^{\mu}\gamma_5)f Z^{'}_{\mu}\,,
\end{equation}
where $V_\chi \equiv  q_\chi g_\chi$ with $q_\chi$ and $g_\chi$ representing the $U(1)'$ charge and gauge coupling respectively. In this case, the spin-independent DM-nucleon scattering cross section in the non-relativistic limit ($(\chi^\dagger \overleftrightarrow{\partial_\mu}\chi)\approx (\chi^\dagger \overleftrightarrow{\partial_0}\chi)\approx(2m_\chi)$) has the same form as Eq.~(\ref{sigmaSI}). 
However, the annihilation cross section $\chi^\dagger \chi\rightarrow \bar{f}f$ has a different form compared to the fermionic DM case. In particular, the derivative coupling leads to a characteristic velocity suppression of the annihilation cross section in the non-relativistic limit. Thus, despite the similarity in the scattering cross section, there will be differences in the final relic abundance and the resulting parameter space for scalar {\it vs.} fermionic DM in this simple heavy vector model.

The annihilation cross section for $\chi^\dagger \chi\rightarrow \bar{f}f$ for the pure vector interaction is given by \cite{Yu:2011by}
\beq
\sigma_{\chi^{\dagger}\chi\rightarrow \overline{f}f}(s) v_{\rm rel} = N_c^f V_f^2V_\chi^2 \frac{\sqrt{1-\frac{4m_f^2}{s}}}{6\pi s}\frac{(s-4m_\chi^2)(s+2m_f^2)}{(s-M_{Z'}^2)^2+\Gamma_{Z'}^2M_{Z'}^2},
\label{scalarDMannCS}
\eeq
 In the non-relativistic limit, velocity suppression of the annihilation cross section relative to that of the Dirac fermion arises from the factor $(s-4 m_\chi^2) = 4 p_\chi^2$. Unlike the Dirac fermion case, we do not consider pure axial vector coupling for scalar DM since the leading order contribution to the spin-dependent scattering cross section in this case will vanish. As a result, we exclusively consider fermion DM for spin-dependent scattering.

The relic density from thermal freeze-out is determined by
the integration of the Boltzmann equation for the dark matter density, $n$
\beq
{\dot n} + 3 H n = - \langle \sigma v_{\rm rel} \rangle (n^2 - n_0^2) \, ,
\eeq
where $n_0$ is the equilibrium dark matter density and $\langle \sigma v_{\rm rel} \rangle$ is the thermally averaged annihilation cross section. The latter is given by
\beq
\langle \sigma v_{\rm rel} \rangle = \frac{1}{n_0^2} \int d^3p_1 d^3p_2 f(E_1) f(E_2) \sigma v_{\rm rel}\, ,
\eeq
where 
\beq
f(E_i) = \frac{g_i}{(2\pi)^3} \left[\exp(E_i/T) \pm 1 \right]^{-1} 
\eeq
and $g_i$ is the number of initial state degrees of freedom.

\begin{figure}[!ht]
\centering
\vskip .2in
\includegraphics[width=4.3in]{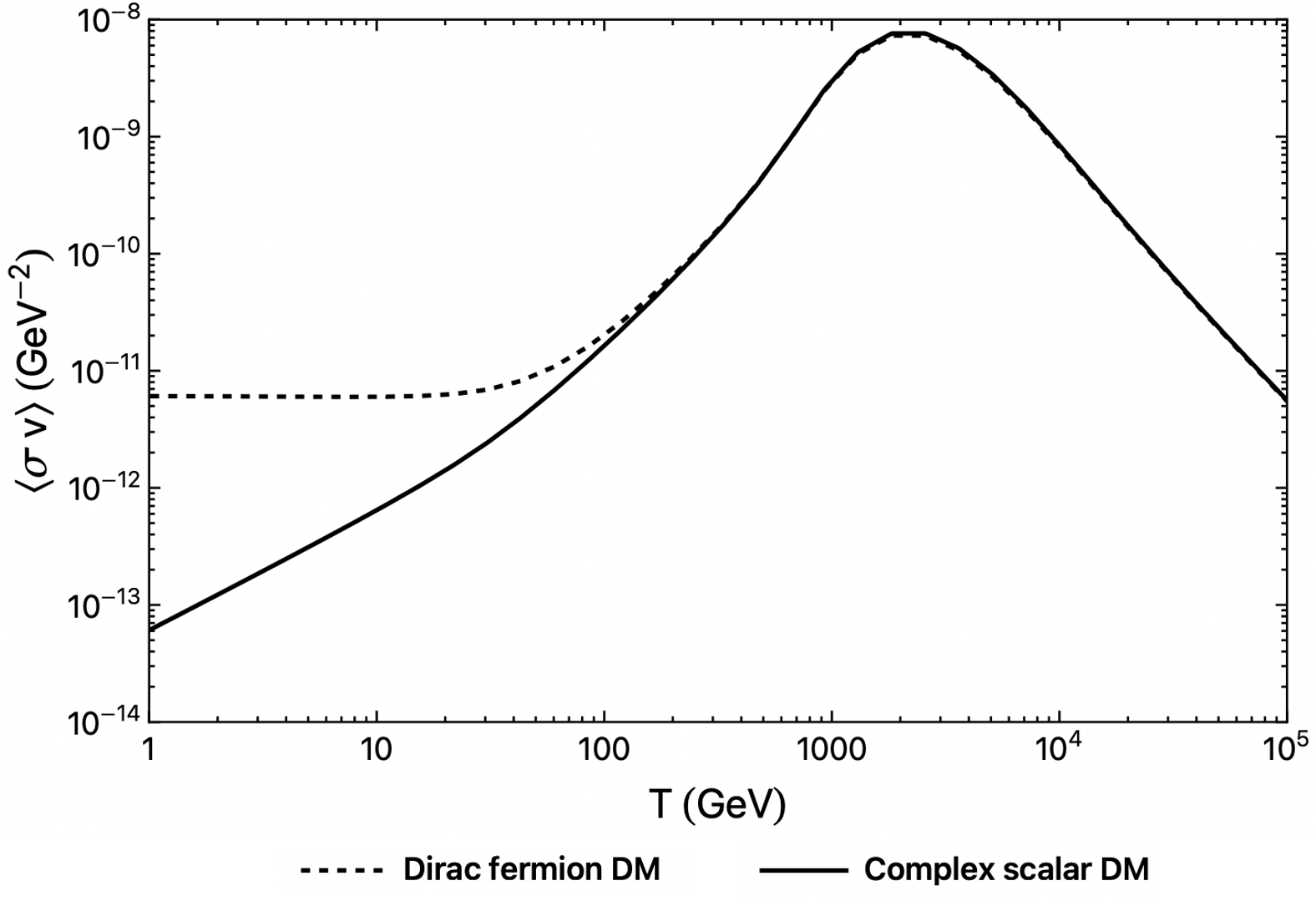}
\caption{\em \small Comparison of the thermally averaged annihilation cross sections for complex scalar DM (solid) {\it vs.} Dirac fermion DM (dashed) via the heavy $Z'$ portal interaction for $m_{\chi}=100$~GeV, $M_{Z'}=10^4$~GeV and $V_f=V_{\chi}=1$. The $p$-wave suppression for scalar DM in the low temperature regime is evident.}
\label{Fig:ScalarVsFermion}
\end{figure}

We plot the thermally averaged annihilation cross sections as a function of temperature for Dirac and scalar DM candidates in 
Fig.~\ref{Fig:ScalarVsFermion}
for $V_{\chi}=V_f=1$, $m_{\chi}=100$~GeV, and $M_{Z'}=10^4$~GeV. 
The $p$-wave suppression at low $T$ is 
clearly visible for the complex scalar (solid) relative to the Dirac 
fermion (dashed). In contrast, the 
thermally averaged cross section for the
Dirac fermion is equal to that of the scalar in the 
high temperature regime. 

\subsection{Constraint from reheating}
The viability of UFO dark matter depends on details of the reheating period. Indeed, 
relativistic decoupling is only possible when interactions between the visible sector and dark matter are no longer efficient enough to counteract the 
dilution due to the expansion of the Universe, governed by the Hubble parameter $H$. This Hubble rate depends on the nature of the field that dominates 
the Universe during reheating. In all the results presented in this paper, we assume that the coherent oscillations of 
the inflaton field, $\phi$, with a quadratic potential $V(\phi)\propto \phi^2$, 
dominate the expansion rate during the reheating phase. Note that while we treat reheating as a non-instantaneous process in this work, we assume instantaneous thermalization after inflation.

On the other hand, interactions between the visible sector and the dark sector 
are governed by the interaction rate $\Gamma \sim n\langle \sigma v \rangle$, where 
the number density $n\sim T^3$ in the relativistic regime and $T$ is the temperature of the standard bath. The evolution of $T$ depends both on the expansion of the Universe (and 
therefore on $H$), and on the efficiency of energy transfer from the inflaton 
to the Standard Model degrees of freedom. In this paper, we assume that this energy 
transfer is a result of inflaton decay which leads to the scaling $T \propto a^{-\frac{3}{8}}$ \cite{Scherrer:1984fd,Giudice:2000ex,GKMO1,GKMO2} where $a$ is the cosmological scale factor. This is in contrast to the typical scaling during adiabatic expansion where $T \propto a^{-1}$.

It is therefore relatively straightforward to understand the a priori complex 
connections that link direct detection to the UFO decoupling regime. Indeed, a 
relatively heavy mediator $Z'$ implies a rather weak coupling between dark 
matter and the Standard Model (leading to small values for $\sigma v$ and $\sigma_{\chi N}$), resulting in early decoupling during the reheating phase, which favors relativistic decoupling. Conversely, a lighter mediator allows for a 
longer period of thermal equilibrium, leading to late decoupling at $T\simeq m_\chi$, 
and therefore to non-relativistic freeze-out.

Note also that  UFO does not necessarily imply the complete halt 
of dark matter production after decoupling, but rather a strongly reduced production compared to particles that remain in thermal equilibrium, much like the case in freeze-in. Indeed, 
Standard Model particles sourced by the inflaton have a comoving density $\simeq n_\gamma \times a^3$ that grows much faster than the comoving density of dark matter after relativistic decoupling, which is 
sourced by the same thermal bath but with a reduced production rate $\propto n_\gamma \langle \sigma v \rangle$. As a consequence, during reheating, the 
ratio $\frac{n_\chi}{n_\gamma}$ decreases until the Universe becomes radiation-dominated, at which point both ${n_\chi}$ and ${n_\gamma}$ scale as $a^{-3}$ and their ratio becomes fixed.

\section{Relic Densities: The WIMP, UFO, and FIMP regimes}
\label{relic}

Using the thermally averaged cross section, we can determine
the relic density from freeze-in  or freeze-out for any choice of $m_\chi, M_Z'$, and $\Trh$. For certain parameter choices, freeze-out may occur while $\chi$ is non-relativistic (WIMP-like freeze-out)
in which case the relic density is essentially fixed at freeze-out.
For other parameter choices,  freeze-out occurs while $\chi$ is relativistic (UFO) and the final relic density may be supplemented by post-freeze-out production. In this case it is necessary to use the reverse cross section (which is also used for freeze-in) for $f {\bar f} \to \chi \chi$ integrated down to $T \simeq m_\chi$. For still other parameter choices, $\chi$,
may never enter into equilibrium and the relic density is produced solely through freeze-in. These possibilities for fermionic dark matter were discussed in detail in \cite{Henrich:2025pca}. Here, we briefly review the result and also include results for scalar dark matter.

\begin{figure*}[!ht]
\centering
\vskip .2in
\includegraphics[width=\textwidth]{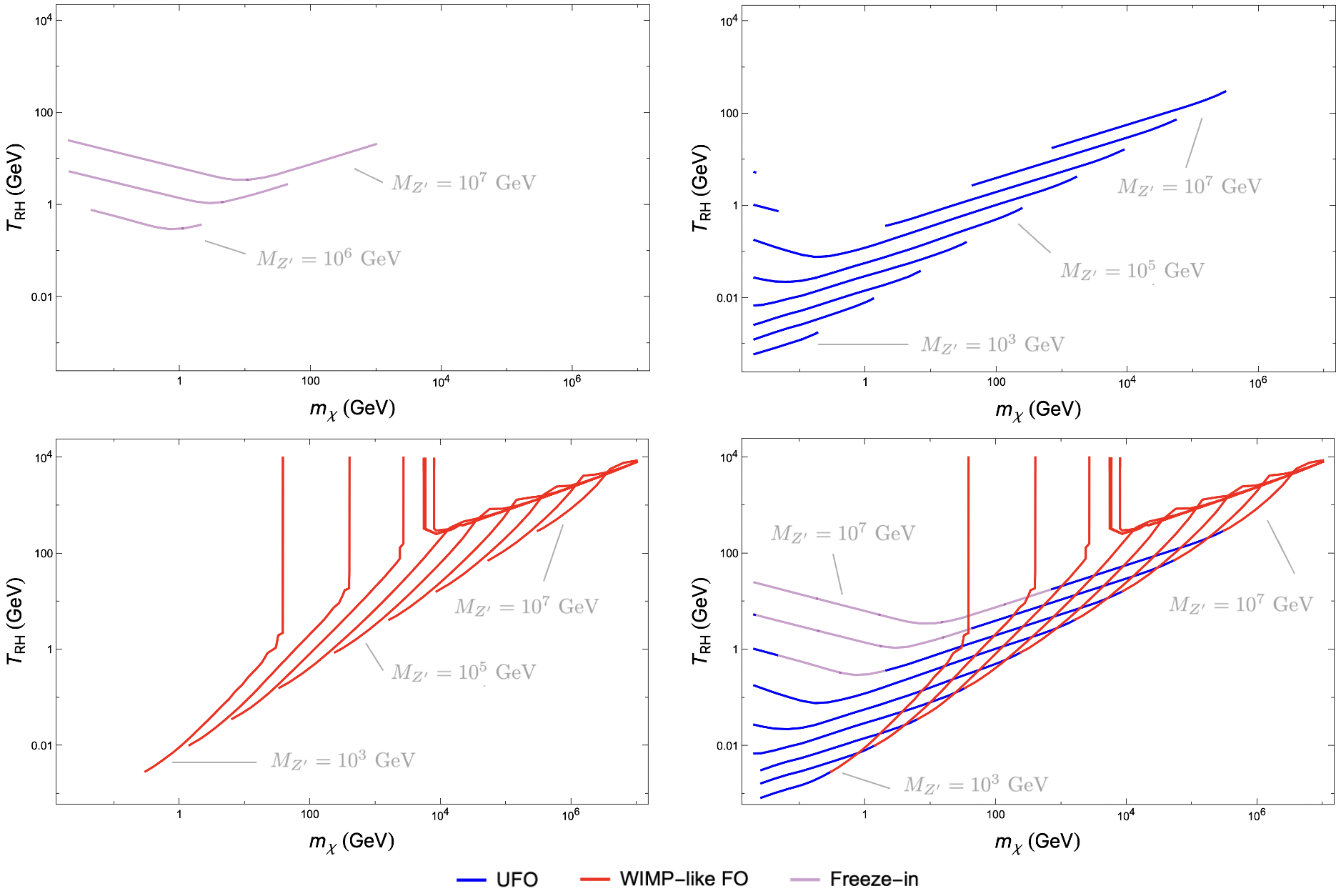}
\caption{\em \small $(m_\chi,T_{\rm RH})$ planes illustrating the allowed parameter space for freeze-in, UFO, and WIMP-like FO for Dirac fermion DM production during reheating. Each contour corresponds to a fixed choice of $M_{Z'}$ ranging from $10^3$~GeV to $10^7$~GeV. The top left (right) panel depicts the freeze-in (UFO) parameter space. The bottom left depicts the WIMP-like FO space, and the bottom right panel is an overlay of all three regimes. The degeneracy between UFO and WIMP-like FO can be observed directly in the bottom right panel at the intersection points between the blue and red curves.}
\label{Fig:UFOWIMPdegenFermionDM}
\end{figure*}

To illustrate the available parameter space, we plot the
$(m_\chi,T_{\rm RH})$ parameter space for several values of $M_{Z'} \le 10^7$~GeV in Figure \ref{Fig:UFOWIMPdegenFermionDM}. The contours in each panel show the value of $\Trh$ for a given $m_\chi$ and $M_{Z'}$ necessary to obtain $\Omega_\chi h^2 = 0.12$.  In the upper
left panel, we show the viable parameter space where DM is produced via freeze-in.  For these values of $m_\chi, M_{Z'}$ and $\Trh$, the thermal production rate cannot compete with the
Hubble rate. In this case, the dark matter never thermalizes. 
{\it However, for lower $M_{Z'}$ and larger DM masses, equilibrium is achieved early on and the dark matter density is determined by UFO (and post-freeze-out production)}, as can be seen in the top right panel of the figure. As we will see, the mass $m_\chi$ must increase with $\Trh$ to ensure a constant relic density in the UFO regime.
Notice that the slope changes for $m_\chi \sim \Trh$. This corresponds 
to a change in the peak production temperature. In particular, the larger of $m_\chi$ and $\Trh$ is approximately where the relevant DM production will peak in the calculation of the relic density. Thus, for all $m_\chi < T_{\rm RH}$, the production peaks near $T_{\rm RH}$. 
For even larger masses $m_\chi$, decoupling occurs when dark matter is non-relativistic, and the freeze-out physics is similar to that of ordinary WIMP-like dark matter. In this regime, the interaction strength is large enough that the DM remains in equilibrium until its equilibrium density begins to be Boltzmann suppressed (unlike UFO, where decoupling occurs prior to the temperatures at which the equilibrium density becomes Boltzmann suppressed). For WIMP-like FO, the annihilation term in the Boltzmann equation is largely what dictates the final abundance, whereas for UFO, it is instead usually the (post-freeze-out) production term which drives the final abundance. \footnote{For IR UFO, which we consider here for $n=2$ and a quadratic inflaton potential (about the minimum), it is indeed the production term in the Boltzmann equation which primarily sets the final abundance. However, for UV UFO (for instance for $n=2$ and a quartic inflaton potential) the final relic density is instead set primarily by the abundance at freeze-out at the UV scale. In this case, the annihilations are very important as well. See \cite{Henrich:2025gsd} for more details regarding this distinction.}

For the present study, we focus on $m_\chi > T_{\rm RH}$  since this region contains DM masses in the range of interest for direct detection ($100 \text{ MeV}\lesssim m_\chi\lesssim 10 \text{ TeV}$) along with 
sufficiently small mediator masses to produce cross sections large enough to be 
detectable. Note that for  $m_\chi \gg T_{\rm RH}$, dark matter production ends before reheating is complete. 
In this region of parameter space, UFO is frequently operative when determining the relic density. This part of the parameter space is shown in the upper right panel of Fig.~\ref{Fig:UFOWIMPdegenFermionDM}. 
When $m_\chi < T_{\rm RH}$, to obtain the correct relic density, we require mediators which are sufficiently heavy that the scattering cross section lies 
exclusively beneath the neutrino fog \cite{OHare:2021utq} and thus currently undetectable by standard terrestrial 
experiments. The parameter space compatible with WIMP-like FO during reheating is shown in the lower left panel of Fig.~\ref{Fig:UFOWIMPdegenFermionDM}. The change in slope for these contours occurs when $m_\chi \simeq M_Z'$.  The lower right panel shows the overlay 
of all three production regimes. For further details, see \cite{Henrich:2025pca}.

The natural boundary between UFO and WIMP-like FO for dark matter produced during the reheating era can be defined by the freeze-out  temperature corresponding to the maximum co-moving equilibrium number density of the dark matter \cite{Henrich:2025pca}. For an inflaton potential which is quadratic about the minimum, the peak equilibrium number density (for $m_\chi \gg T_{\rm RH}$) occurs at $T\simeq\frac{2}{13}m_\chi$. Thus, if $T_{\rm FO}>\frac{2}{13}m_\chi$, (ultra)relativistic freeze-out occurs, while if $T_{\rm FO}<\frac{2}{13}m_\chi$, non-relativistic freeze-out during reheating will be responsible for the final relic density. Strictly speaking, the proper definition of ultrarelativistic freeze-out might correspond to about $T_{\rm FO} \gtrsim 3 m_\chi$ while $3m_\chi \gtrsim T_{\rm FO} \gtrsim \frac{2}{13}m_\chi$ might more accurately be called semi-relativistic freeze-out. For simplicity, we will classify all $T_{\rm FO}>\frac{2}{13}m_\chi$ as UFO. For further details regarding the naturalness of this definition, see \cite{Henrich:2025pca}. Note that the above distinction between relativistic and non-relativistic FO is suitable when $T_{\rm max}\gg m_\chi \gg T_{\rm RH}$ which holds for the majority of our parameter space. If $m_\chi$ approaches $T_{\rm RH}$, this distinction will be modified. If $m_\chi >T_{\rm max}$ (which we do not consider here since $T_{\rm max}$ is very large for the high-scale inflation models we have in mind), then production would proceed via Boltzmann-suppressed freeze-in, which was first studied in detail in \cite{Giudice:2000ex} (although the "freeze-in" terminology was not yet in use). 

Interestingly, the authors in \cite{Giudice:2000ex} also considered the case of relativistic decoupling in their Section D, but they concluded that relativistic freeze-out during reheating was impossible. This seems to be because the authors used the condition $n_\chi(T_{\rm RH})<n_{\chi, \rm eq}(T_{\rm RH})$ for determining whether the DM ever reached equilibrium at some point during reheating. However, this condition is too strong, since it is generally possible for $\chi$ to equilibrate at much higher temperatures (e.g. near $T\approx M_{Z'}\gg T_{\rm RH}$) before decoupling relativistically and then later leading to $n_\chi(T_{\rm RH})<n_{\chi, \rm eq}(T_{\rm RH})$ despite the fact that $\chi$ equilibrated early. This scenario is precisely what we are studying in this work.

During reheating, the ratio of the density
of relativistic particles to the density of UFO dark matter, $\propto n_\gamma/n_\chi$, increases after FO, up to $\Trh$ (if $m_\chi< \Trh$), or up to $\approx m_\chi$ otherwise. 
As noted earlier, 
for $n=2$ and a quadratic inflaton potential, the UFO production of DM is IR, thus leading to peak production near $T\simeq \text{max}(T_{\rm RH},m_\chi)$.
For $m_\chi > \Trh$, the current fraction of critical density in $\chi$ can be expressed in terms of the number density, $n_\chi$ evaluated at reheating, $m_\chi$, and the reheating temperature \cite{mybook}
\beq
\Omega_{\chi}h^2\simeq 5.88 \times 10^6 \left( \frac{427}{4 g_{\rm RH}} \right) \, \left(\frac{n_{\chi}(a_{\rm RH})}{T_{\rm RH}^3}\right)\left(\frac{m_\chi}{1~\rm GeV}\right)\,,
\label{Eq:omegageneric}
\eeq
where $g_{\rm RH}$ is the number of relativistic degrees of freedom at $\Trh$. 
The comoving number density of UFO dark matter was computed in \cite{Henrich:2025gsd} where it was found that\footnote{Here we include an additional factor of 4 in the prefactor (removing the spin averaging in \cite{Henrich:2025gsd} which is partially compensated for by decreasing $\Sigma_{\rm tot}$ by a factor of 2. } 
\beq
Y_\chi(a_{\rm m}) \approx Y_\chi(a_{\rm RH}) =  Y_\chi(a_{\rm FO})+
9.1 \times 10^{-3} \Sigma_{\rm tot}
\frac{T_{\rm RH}^{6}M_{P}}{M_{Z'}^{4}} a_{\rm RH}^\frac32 \left[
a_{\rm m}^\frac32 - a_{\rm FO}^\frac32\right]\,,
\label{Eq:yrhUFOmGeqTRH}
\eeq
where $Y_\chi =  n_\chi a^3$, and $a_{\rm RH}, a_{\rm m}$, and $a_{\rm FO}$ are the values of the scale factor at reheating, when $T = m_\chi$, and at freeze-out respectively and we have taken $g^*(a_m)=427/4$. The factor $\Sigma_{\rm tot}$ is given by
\beq
\Sigma_{\rm tot} = \sum_f \frac12 \left(V_f^2 V_\chi^2 + V_f^2 A_\chi^2 + A_f^2 V_\chi^2 + A_f^2 A_\chi^2 \right) + 6 \left[ A_\nu V_\nu \left( A_\chi^2 + V_\chi^2 \right) \right]
\label{newsigmatot}
\eeq
so that $\Sigma_{\rm tot} = 45/2$ for $V_f = V_\chi =1$, $A_f=A_\chi=0$, and the sum is over all 45 Standard Model fermions.\footnote{The 45 states include 24 left-handed states and 21 right handed states. The absence of right handed neutrinos gives rise to the last term in (\ref{newsigmatot}) which vanishes for pure vector or axial vector interactions.}
Note that the factor of 45/2 assumes that $m_\chi > m_t$ and thus for lower $m_\chi$, the number of fermions we sum over must be adjusted accordingly. This is always done in all of our numerical calculations.
For $a_{\rm FO} \ll a_{\rm m}$, $Y_\chi(a_{\rm FO})$ is typically small compared to the post-freeze-out contribution\footnote{$Y_\chi(a_{\rm FO})$ is typically much smaller than the post-freeze-out contribution for the IR UFO scenario considered in this work. For the UV UFO scenario, this term cannot be neglected.}, and using $(a_{\rm m}/a_{\rm RH})^3 = (\Trh/m_\chi)^8$, 
we arrive at\footnote{Note that in our numerical work, we also take into account the changes in the numbers of degrees of freedom between $T=m_\chi$ and $\Trh$. }
\beq
\frac{\Omega_\chi h^2}{0.12}
\simeq 1.1 \times 10^{12} \Sigma_{\rm tot} \left( \frac{427}{4 g_{\rm RH}} \right) \left(\frac{\Trh}{1~\rm GeV}\right)^7
\left(\frac{1~\rm TeV}{M_{Z'}}\right)^4
\left(\frac{1~\rm GeV}{m_\chi}\right)^3\,.
\label{Eq:omegaapprox}
\eeq
For fixed $M_{Z'}$, setting $\Omega_\chi h^2 = 0.12$, provides a
relation between $\Trh$ and $m_\chi$, which we see from 
Eq.~(\ref{Eq:omegaapprox}) is
 $\Trh \propto m_\chi^{3/7}$ as is found in the upper right panel of Fig.~\ref{Fig:UFOWIMPdegenFermionDM} for the positive sloped portion of the blue curves. 
 We note that the numerical expression in Eq.~(\ref{Eq:omegaapprox})
 is an approximation which assumes that production of $\chi$ ceases at $a_{\rm m}$. However as discussed in \cite{Henrich:2025gsd}, some production continues for $T < m_\chi$ potentially increasing the estimate in Eq.~(\ref{Eq:omegaapprox}) by a factor $\mathcal{O}(50-250)$. 
 The larger factors occur when the UFO/WIMP transition is approached. In that regime, the value of $Y_\chi (a_{\rm FO}) $ cannot be neglected and is not included in the analytical estimate (\ref{Eq:omegaapprox}). 
 However, this affects the estimate of $\Trh$ only by a factor of $\mathcal{O}(2)$, due to the steep dependence of $\Omega_\chi$ on $\Trh$. 

For the regime in which the DM undergoes WIMP-like non-relativistic freeze-out (red contours in Fig.~\ref{Fig:UFOWIMPdegenFermionDM}), it is generally possible for there to be parameter values consistent with $\Omega_\chi h^2=0.12$ for DM production during either reheating or radiation domination (RD). For the $Z'$ masses we consider here paired with strong couplings (i.e. $V_{f}=V_{\chi}=1$), the only RD solutions in our DM mass range correspond to non-relativistic freeze-out, rather than UFO or FI. While the majority of our relevant parameter space corresponds to DM production during the reheating era, there is a subset of the non-relativistic freeze-out parameter space for sufficiently small $M_{Z'}$ which is compatible with RD solutions. In Fig.~\ref{Fig:UFOWIMPdegenFermionDM}, the RD solutions are shown as vertical portions of the red contours for $10^3 \leq M_{Z'} \leq 10^4$~GeV. The segments are vertical because the value of the reheating temperature becomes irrelevant if the production occurs during RD, for sufficiently large $T_{\rm RH}$. It is interesting to note that had we taken $M_{Z'} = M_Z \simeq 90$~GeV, the corresponding vertical vertical curve would appear at $m_\chi \simeq 4$~GeV as in the case of a heavy neutral lepton \cite{hut}.  For $M_{Z'}\gtrsim 10^4$~GeV, the DM-SM interactions are too weak to permit RD solutions, as freeze-out occurs during reheating.

\begin{figure}[!ht]
\centering
\vskip .2in
\includegraphics[width=4.3in]{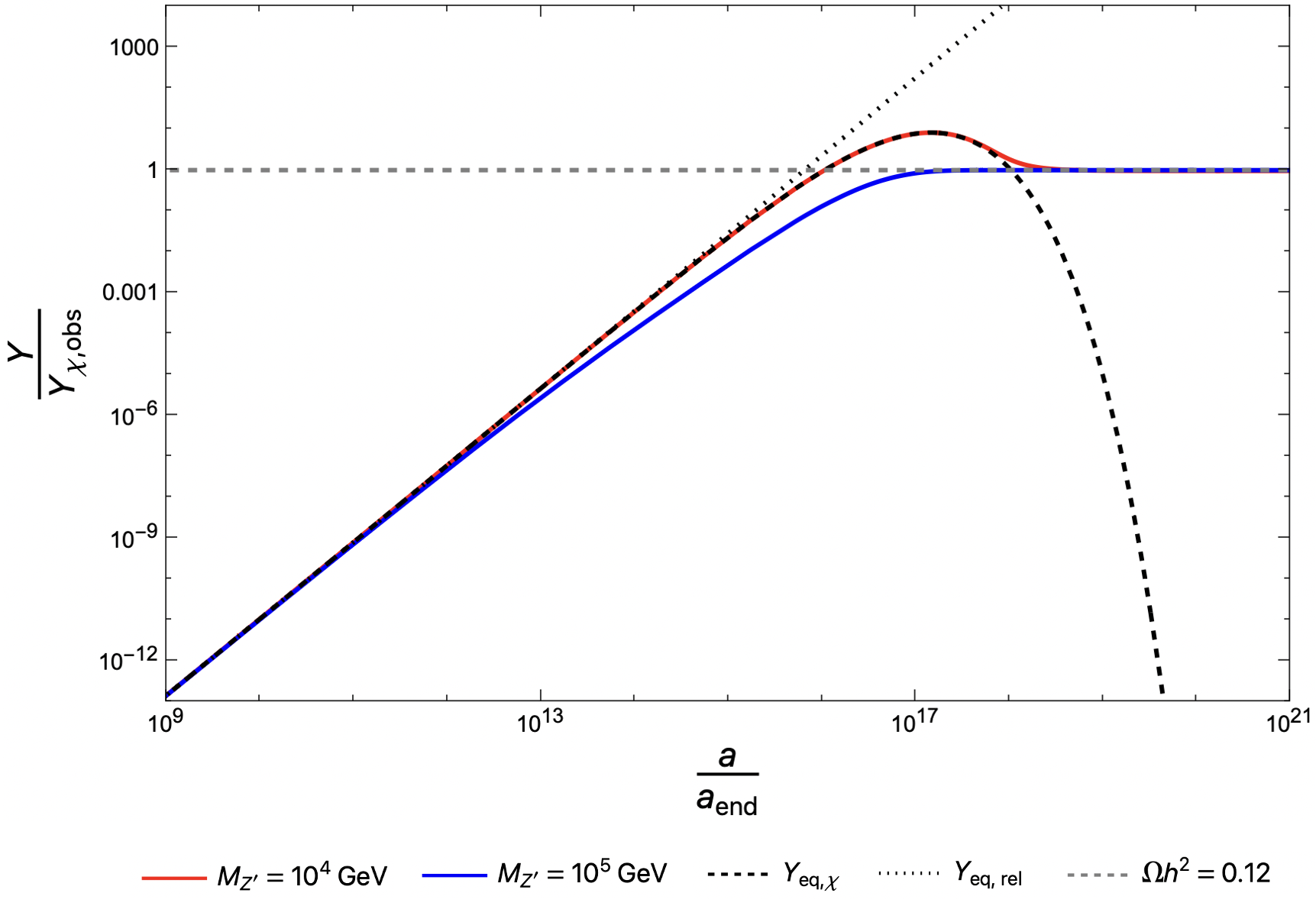}
\caption{\em \small Evolution of the co-moving abundance of dark matter, $Y_\chi=n_\chi a^3$, illustrating the possible degeneracy between UFO and non-relativistic WIMP-like FO for fixed $m_\chi$ and $T_{\rm RH}$. Two different values of $M_{Z'}$ can yield the same abundance. The above uses $m_\chi = 55$~GeV, $T_{\rm RH}=450$~MeV and $M_{Z'}=10^4$~GeV (red) and $M_{Z'}=10^5$~GeV (blue). The black dashed (dotted) lines show the co-moving DM equilibrium abundance (co-moving equilibrium abundance of electrons). The horizontal gray dashed line corresponds to $\Omega h^2=0.12$. The x-axis is normalized to $a_{\rm end}$, which is the scale factor at the end of inflation.}
\label{Fig:UFO_WIMPdegen}
\end{figure}

Note that in some areas, there are intersections of UFO and WIMP contours which 
represent a degeneracy which can be understood as follows. 
For heavy 
portal models where DM production occurs during reheating (or early matter domination),
there is a possible degeneracy between WIMP-like non-relativistic freeze-out and 
(ultra)relativistic freeze-out for a fixed DM mass and reheating temperature. 
This can be 
seen in Fig.~\ref{Fig:UFO_WIMPdegen}, 
which depicts the evolution of the co-moving DM 
abundance $Y_\chi=n_\chi a^3$ for two choices of $M_{Z'}$ and identical $m_\chi$ and $T_{\rm RH}$. 
We see that the correct final abundance ($\Omega_\chi h^2=0.12$) is 
recovered in both cases, one via WIMP-like FO (red) and one via UFO (blue).  The black dashed curve shows the equilibrium comoving density for a massive relic undergoing annihilations. The black dotted line shows the comoving density for a relativistic particle in equilibrium. In this plot, we have fixed the values of $m_\chi$ and $T_{\rm RH}$, but we have taken  
two choices of $M_{Z'}$. 
For the larger value of  $M_{Z'} = 10^5$~GeV,  annihilations do not play a significant role and $\chi$ freezes out while relativistic (UFO) and we see a departure from the equilibrium abundance after freeze-out, due its dilution with respect to the thermal bath. However, freeze-in-like production continues, and the comoving abundance increases until $T\lesssim m_\chi$. Subsequently, $Y_\chi$ becomes constant. 
For the smaller value of $M_{Z'}= 10^4$~GeV 
the DM remains in equilibrium long enough that $\chi$ becomes non-relativistic and annihilations reduce the abundance exponentially until freeze-out.  After freeze-out, the comoving density shown by the red curve becomes constant, with the same value as the case with higher $M_Z'$.

This captures the 
essence of the degeneracy between WIMP-like FO and UFO. 
Note that the fact that decoupling occurs {\it during} reheating is fundamental to this type of degeneracy. For high reheating temperatures, relativistic freeze-out leads to a strong constraint on the DM mass $m_\chi < \mathcal{O}(100)$~eV \cite{Henrich:2025gsd}. Indeed, it is the continuous production of entropy following the UFO period that allows for the effective dilution of dark matter and allows for higher masses. 
However, the fact that this  degeneracy is possible does not mean that it is inevitable.
Indeed, for certain regions of parameter space WIMP-like FO is prohibited since it would require couplings that violate 
perturbativity to ensure the right relic abundance; as a result, these regions exhibit no degeneracy. Conversely, if the DM mass is large enough (e.g. if $m_\chi> M_{Z'}$), then UFO is not possible and again there will be no degeneracy. 

\begin{figure*}[!ht]
\centering
\vskip .2in
\includegraphics[width=\textwidth]{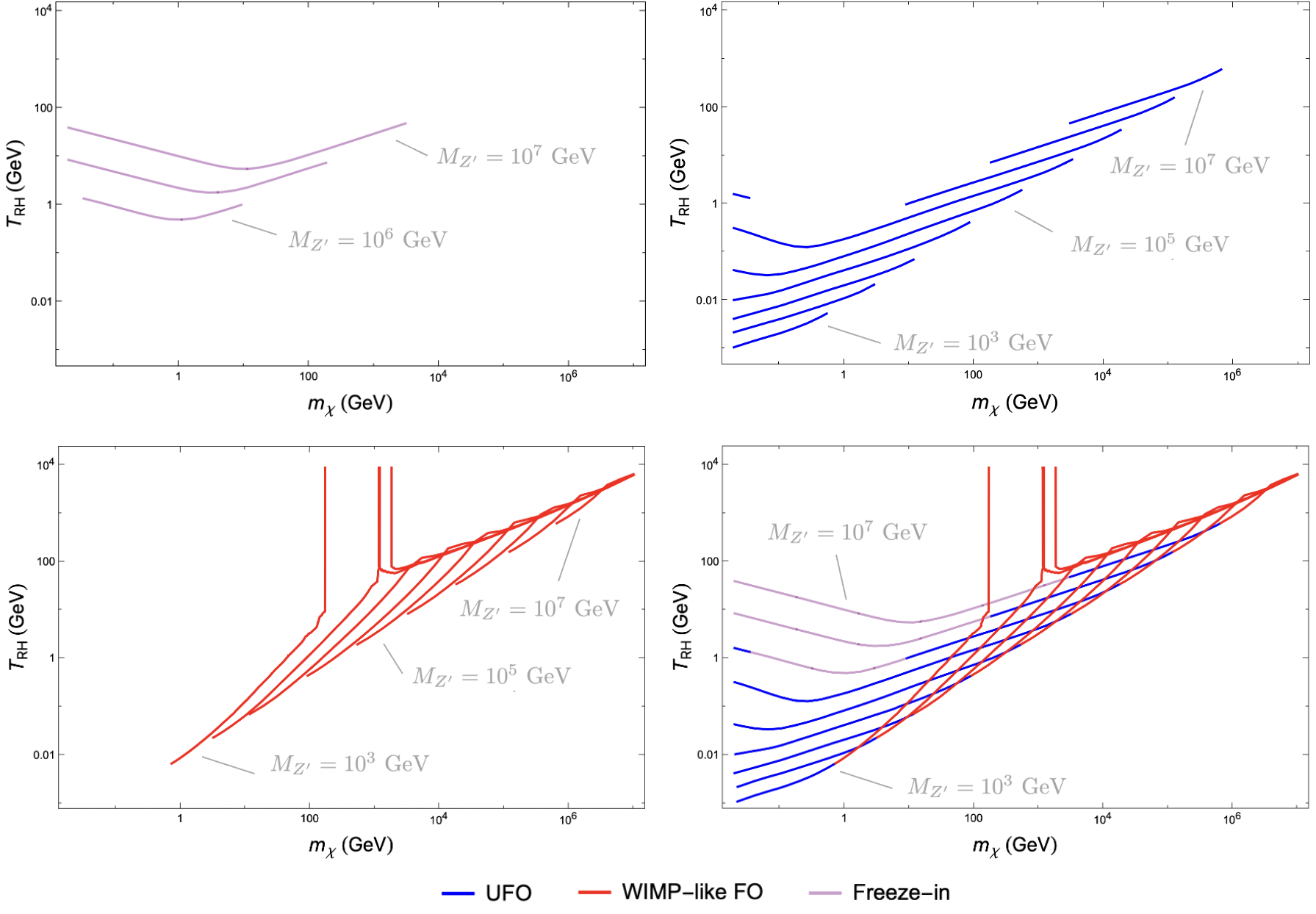}
\caption{\em \small $(m_\chi,T_{\rm RH})$ planes illustrating the allowed parameter space for freeze-in, UFO, and WIMP-like FO for scalar DM production during reheating. Each contour corresponds to a fixed choice of $M_{Z'}$ ranging from $10^3$~GeV to $10^7$~GeV. The top left (right) panel depicts the freeze-in (UFO) parameter space. The bottom left depicts the WIMP-like FO space, and the bottom right panel is an overlay of all three regimes. The degeneracy between UFO and WIMP-like FO can be observed directly in the bottom right panel at the intersection points between the blue and red curves.}
\label{Fig:UFOWIMPdegenScalarDM}
\end{figure*}

The results for the relic density for a scalar DM candidate are qualitatively similar and the $(m_\chi,\Trh)$ planes for scalar DM are shown in Fig.~\ref{Fig:UFOWIMPdegenScalarDM}. As in Fig.~\ref{Fig:UFOWIMPdegenFermionDM}, we display separately the regions where the relic density is determined by freeze-in (upper left), UFO (upper right) and non-relativistic freeze-out (lower left). The full set of curves are combined in the lower right panel of Fig.~\ref{Fig:UFOWIMPdegenScalarDM} for scalar DM where we again see parameter values yielding the same relic density through either UFO or non-relativistic freeze-out. As in the fermion DM case above, there is a small subset of parameter space corresponding to RD solutions visible as vertical red contour segments for sufficiently small $M_{Z'}$.

\begin{figure}[!ht]
\centering
\vskip .2in
\includegraphics[width=4.3in]{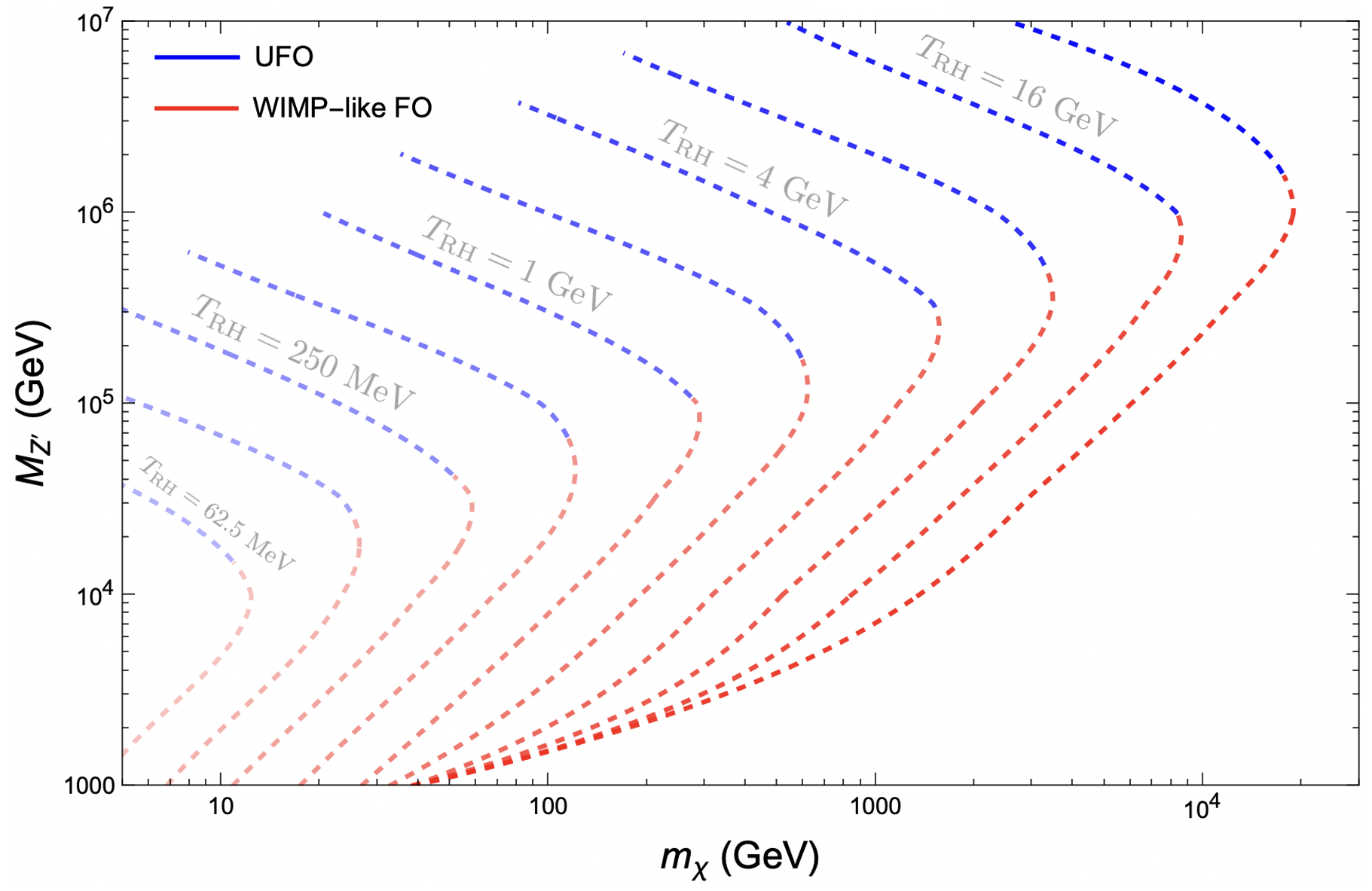}
\caption{\em \small $(m_\chi,M_{Z'})$ plane illustrating the observed relic abundance for Dirac fermion DM for several choices of $T_{\rm RH}$. The blue portions of each curve correspond to UFO while the red portions correspond to WIMP-like FO.
}
\label{Fig:mvsMZ}
\end{figure}

\begin{figure}[!ht]
\centering
\vskip .2in
\includegraphics[width=4.3in]{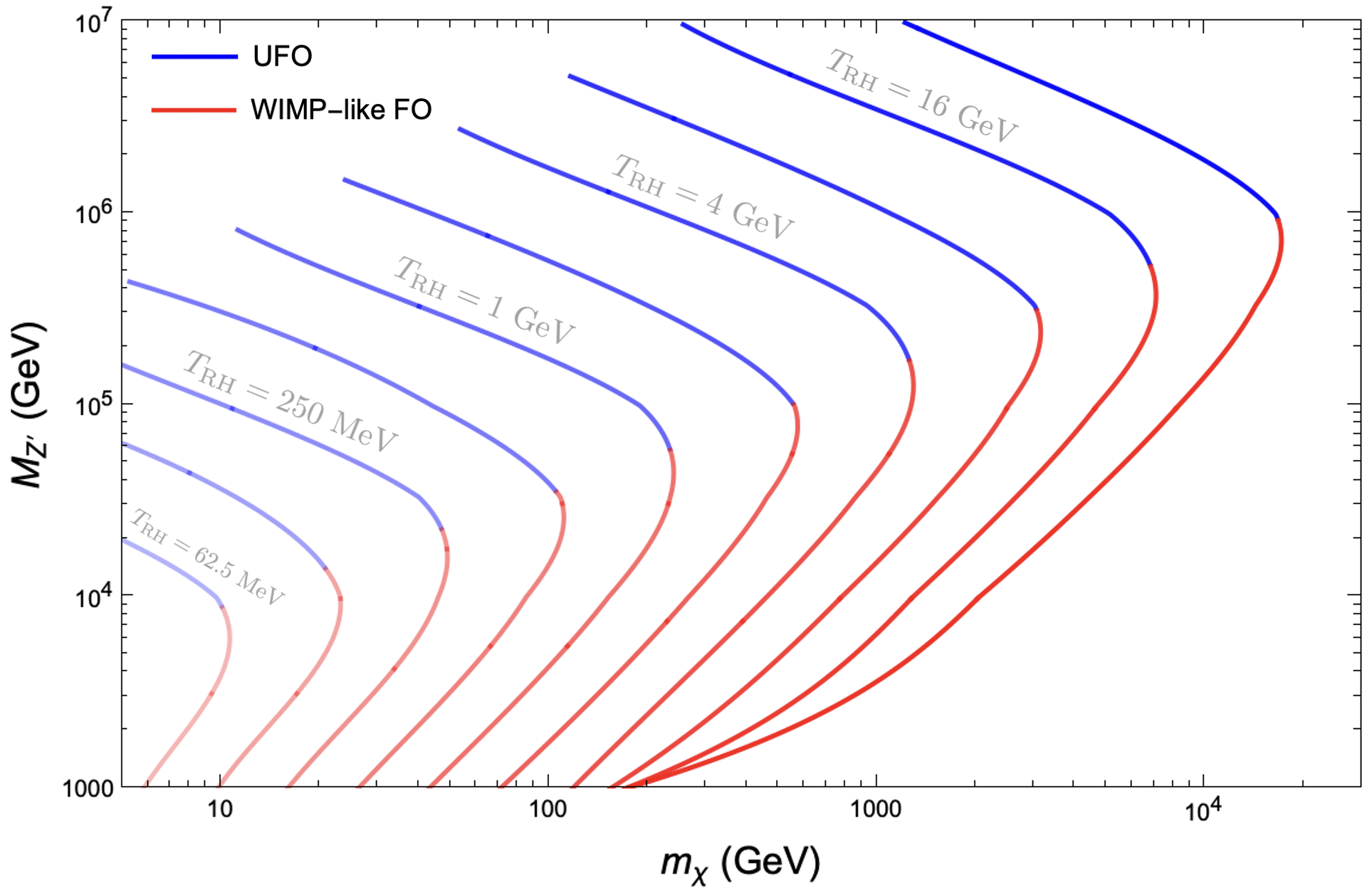}
\caption{\em \small $(m_\chi,M_{Z'})$ plane illustrating the observed relic abundance for scalar DM for several choices of $T_{\rm RH}$. The blue portions of each curve correspond to UFO while the red portions correspond to WIMP-like FO.}
\label{Fig:mvsMZScalar}
\end{figure}

The WIMP/UFO degeneracy is also illustrated in 
Fig.~\ref{Fig:mvsMZ} (Fig.~\ref{Fig:mvsMZScalar}), 
for a Dirac fermion (scalar) DM candidate. 
In these figures we plot contours corresponding to a relic abundance, $\Omega_\chi h^2 = 0.12$ in the ($m_\chi, M_{Z'}$) plane with fixed $T_{\rm RH}$. 
Blue portions of the curves correspond to UFO during reheating while the 
red portions of the curves correspond to WIMP-like FO. %
The degeneracy between the 
two mechanisms is clearly visible: for a given $m_\chi$ (and $\Trh$), the same relic density can be obtained through either  
non-relativistic freeze out in the WIMP-like regime (red) {\it or} in the UFO regime 
(blue). The latter is achieved with a heavier $M_{Z'}$,
as explained above.

\section{Direct detection constraints}
\label{results}

\subsection{Spin-independent scattering}

\begin{figure}[!ht]
\centering
\vskip .2in
\includegraphics[width=4.3in]{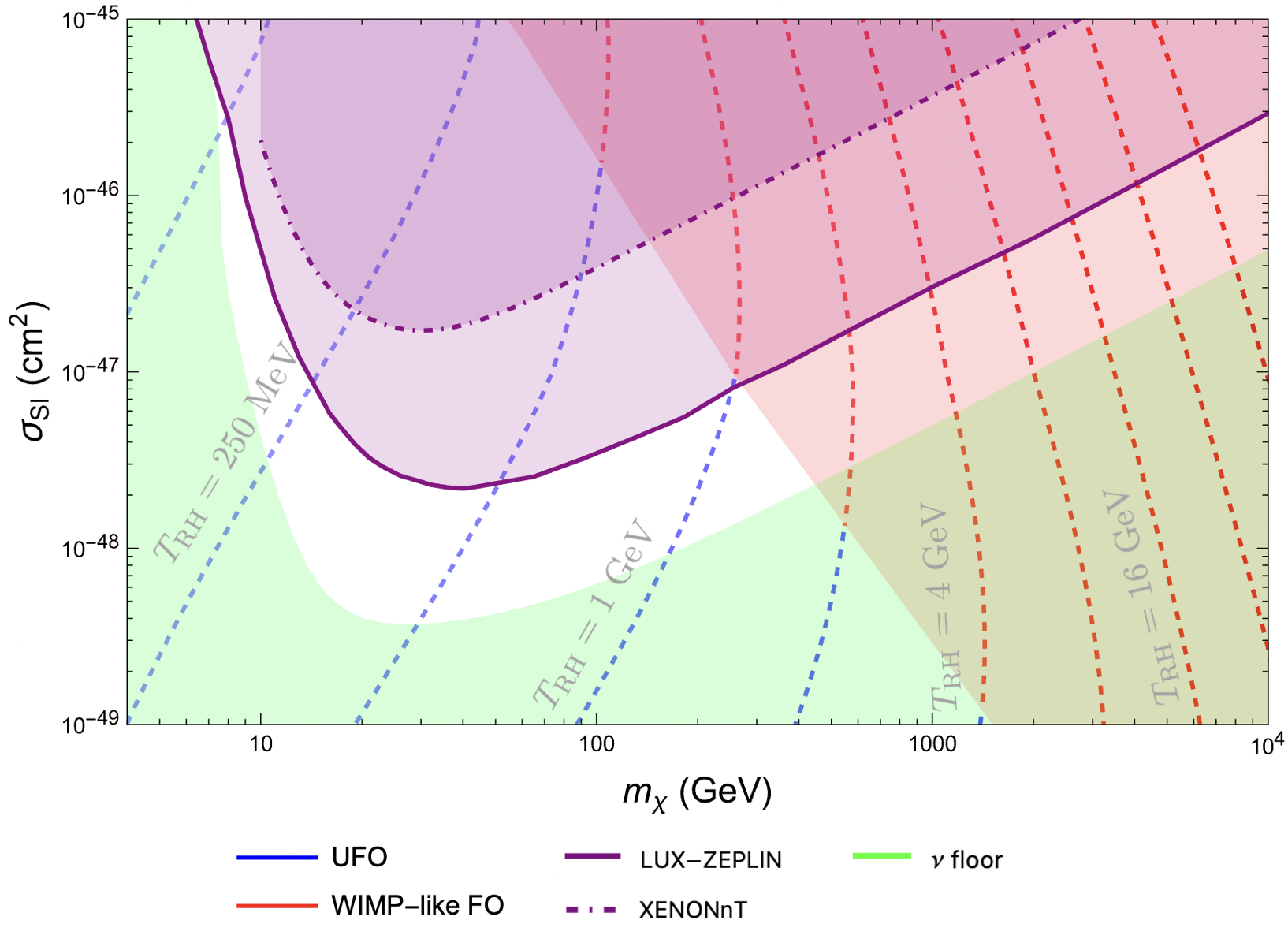}
\caption{\em \small Direct detection limits from XENONnT \cite{XENON:2025vwd} and LUX-ZEPLIN \cite{LZ:2024zvo, LZ:2025igz}
on the spin-independent nuclear scattering cross section for Dirac fermion UFO dark matter. The purple shaded regions are excluded while the white regions correspond to UFO and are allowed. Blue (red) dashed contours correspond to UFO (WIMP) production for fixed values of $T_{\rm RH}$. The red shaded region corresponds to WIMP-like FO during reheating and the green shaded region is the neutrino fog. We have used $V_q=V_\chi=1$. For each contour with fixed $T_{\rm RH}$, the value of $M_{Z'}$ will vary depending on $m_\chi$ to obtain the correct abundance. Typical values of $M_{Z'}$ in the white UFO region are $\mathcal{O}(300)$~TeV.}
\label{Fig:DDlimitsFermion}
\end{figure}

Equipped with the DM-nucleon scattering cross section for the  
vector portal
interactions, along with the parameters consistent with the correct relic abundance, 
we can chart the territory in the $(m_\chi, \sigma_{\chi N})$ plane to determine the experimentally excluded (and accessible) regions.
First, we consider the spin-independent cross section $\sigma_{\rm SI}$. 
For simplicity, we consider exclusively vector portal interactions and universal 
couplings, which corresponds to $V_\chi=V_f=1$ and $A_\chi=A_f=0$. Generalization is straightforward, by rescaling the results.
The $(m_\chi, \sigma_{\chi N})$ plane for a fermionic DM candidate is shown in Fig.~\ref{Fig:DDlimitsFermion}, for several choices of the  reheating temperatures. Along  each contour for a given reheating temperature, the 
mass of the mediator $M_{Z'}$ is determined by the relic density,
and varies from its maximum value at the bottom of the curve (blue UFO regime) to its minimum value at the top of the curve (red WIMP regime).
The $Z'$ mass range considered is $10^3\leq M_{Z'}\leq 10^{7}$~GeV. We also show the current constraints obtained by the XENONnT \cite{XENON:2025vwd} and LUX-ZEPLIN \cite{LZ:2024zvo,LZ:2025igz} collaborations in purple, while the neutrino fog is shown in green.  The transition between the WIMP/UFO regimes is depicted by the shaded  
region on the right where the contours change color from blue (UFO) to red (WIMP).  Clearly the increased sensitivity of direct detection experiments combined with the lack of a positive signal, has narrowed the gap between the 90 \% CL upper limit on the cross section and the neutrino fog, leaving the area between the purple and green shaded regions as viable and not yet explored. 

From Fig.~\ref{Fig:DDlimitsFermion}, we see that ongoing direct detection experiments are already 
probing the UFO parameter space, and have in fact excluded some UFO regions in the DM mass regime between a few GeV and $\sim200$~GeV for fermionic UFO DM. 
Viable UFO 
parameter space above the neutrino fog and below the existing detection 
limits extends to  $m_\chi \simeq 400$~GeV. Furthermore, reheating temperatures below about 2 GeV 
are necessary for direct detection prospects for UFO DM. 
We note that UFO DM is perfectly 
compatible with much larger reheating temperatures, even up to $T_{\rm RH}\approx 10^{14}$~GeV as it was shown in \cite{Henrich:2024rux,Henrich:2025sli,Henrich:2025gsd}. However,  higher reheating temperatures require  
heavier mediator masses to ensure an early relativistic freeze out during reheating, in which case these UFO DM candidates will 
be undetectable by existing experiments as their scattering cross sections fall well below the neutrino fog.

Using the analytical estimate for the relic density in Eq.~(\ref{Eq:omegaapprox}) together with the expression for the scattering cross section in Eq.~(\ref{Eq:sigmasi}), we can get a better understanding of the contours in Fig.~\ref{Fig:DDlimitsFermion}. In the limit $m_\chi \gg m_N$, 
$\mu_{\chi N} \approx m_N$ and for $V_\chi = V_q = 1$, 
Eq.~(\ref{Eq:sigmasi}) gives
\beq
\sigma^{\rm SI}_{\chi N} \simeq \frac{9 m_N^2}{\pi M_Z'^4} \simeq \frac{2.5~\rm GeV^2}{M_{Z'}^4} \,.
\label{Eq:mzpsigmasi}
\eeq
 Combining this expression with Eq.~(\ref{Eq:omegaapprox}), we expect that contours with $\Omega_\chi h^2 = 0.12$ correspond to a
scattering cross section given by
\beq
\sigma^{\rm SI}_{\chi N} \simeq 9.1 \times 10^{-52} {\rm cm}^2 \frac{4g_{\rm RH}}{427\Sigma_{\rm tot}} \left(\frac{1~\rm GeV}{\Trh}\right)^7 \left(\frac{m_\chi}{1~\rm GeV}\right)^3
\label{Eq:sigmaUFO}
\eeq
or $\sigma^{\rm SI}_{\chi N}\propto m_\chi^3$, which is effectively what we observe on the blue lines of Fig.~\ref{Fig:DDlimitsFermion} for fixed $\Trh$ when we take into account the additional production (by a factor of $\mathcal{O}(200)$) for $T< m_\chi$, as noted above.

In contrast to the dependence of the scattering cross section on the DM mass in the UFO regime (as in Eq.~(\ref{Eq:sigmaUFO}))
the dependence is expected to be very different for WIMP-like DM candidates. In the case of a WIMP, non-relativistic 
freeze-out is determined when the WIMP annihilation rate falls below the expansion rate and the relic density is determined by 
the non-relativistic limit of the thermally averaged annihilation cross section. If 
freeze-out occurs {\it after} reheating (as in the common thermal freeze-out scenario) the relic density can be expressed as \cite{mybook}
\beq
\frac{\Omega_\chi h^2}{0.12}
\simeq 1.4 \times 10^{-9}~{\rm GeV}^{-2} \frac{1}{\sqrt{g_{\rm FO}} \langle \sigma v \rangle x_{\rm FO}} \, ,
\label{standfo}
\eeq
where $g_{\rm FO}$ is the number of degrees of freedom when non-relativistic freeze-out occurs, $x_{\rm FO} \equiv T_{\rm FO}/m_\chi$, and $\langle \sigma v \rangle$ is the s-wave annihilation cross section. Note that we have included a factor of 2 for a Dirac $\chi$. Typically, $x_{\rm FO} \sim 1/20$, though its exact value will depend on the annihilation cross section which can be taken from Eq.~(\ref{sannv}).
For fermionic $\chi$, the relic density can be rewritten as
\beq
\frac{\Omega_\chi h^2}{0.12} \simeq 3.8 \times 10^{-10}~{\rm GeV}^{-2} \sqrt{\frac{427}{4 g_{\rm FO}}} \frac{M_{Z'}^4}{m_\chi^2} \, ,
\label{standfo2}
\eeq
 where we have assumed that $m_f \ll m_\chi$ (valid in the upper right of Fig.~\ref{Fig:DDlimitsFermion}) and have summed over all Standard Model fermions (accounting for a factor of 45/2) to obtain the total annihilation cross section.

However, for $\Trh < T_{\rm FO}$, when freeze-out occurs {\it during} reheating (as in the case of interest here), the standard solution for the relic density given by Eq.~(\ref{standfo}) is no longer valid as the expansion rate used to derive (\ref{standfo}) is not driven by the radiation density but rather the inflaton energy density if freeze-out occurs before reheating. 
During reheating, the temperature of the radiation density falls as $T \propto a^{-3/8}$. If we define $q=n_\chi/T^8$, then the Boltzmann equation during reheating can be expressed as 
\beq
\frac{dq}{dx} = \frac83 m_\chi \langle \sigma v \rangle (q^2 - q_0^2) \frac{T^7}{H} \, ,
\eeq
where $q_0$ is the scaled equilibrium density of $\chi$, and the expansion rate can be written as
 \beq
H = \frac{\sqrt{\pi^2 g} T^4}{\sqrt{90}M_P \Trh^2}
 \eeq
For $q_0 \ll q$, this equation is easily solved leading to
\beq
\frac{n_\chi(a_{\rm RH})}{\Trh^3} = \frac{\sqrt{3}}{2} \sqrt{\frac{\pi^2 g_{\rm FO}}{30}} \frac{\Trh^3}{m_\chi^4 \langle \sigma v \rangle x_{\rm FO}^4 M_P} \, ,
\eeq
for a constant (s-wave) cross section. 
This expression can be inserted to into Eq.~(\ref{Eq:omegageneric})
to give 
\bea
\frac{\Omega_\chi h^2}{0.12}  &=&  2.9 \times 10^{-11} ~{\rm GeV}^{-2} \frac{M_{Z'}^4 \Trh^3}{m_\chi^5 x_{\rm FO}^4} \nonumber \\
& \simeq & 1.2 \times 10^{-5} \frac{\Trh^3}{m_\chi^5 \sigma^{\rm SI}_{\chi N}}\,,
\label{wfo}
\eea
 for $g_{\rm RH} = g_{\rm FO} = 427/4$, and $x_{\rm FO} = 1/20$ in the second expression of (\ref{wfo}). We considered a vectorial--type coupling with $V_i=1$, $A_i=0$ (see Eq.~(\ref{sannv})), and used Eq.~(\ref{Eq:mzpsigmasi}) to eliminate $M_{Z'}^4$.
Thus we expect the red contours in Fig.~\ref{Fig:DDlimitsFermion} to be given by
\beq
\sigma^{\rm SI}_{\chi N} \simeq 4.5 \times 10^{-33} {\rm cm}^2 \left(\frac{\Trh}{1{\rm GeV}} \right)^3 \left(\frac{1{\rm GeV}}{m_\chi} \right)^5 \, ,
\eeq
in reasonably good agreement with the more precise numerical evaluation used in Fig.~\ref{Fig:DDlimitsFermion}.

 We would like to emphasize again, the degeneracy between UFO and WIMP-like freeze out. 
For a fixed value of $T_{\rm RH}$, there exists a critical value of $\sigma_{\rm SI}$ for 
which the production process transitions from UFO to WIMP-like FO during reheating. As a result, for a given mass (and $\Trh$), we may find two resulting elastic cross sections associated with $\Omega_\chi h^2 = 0.12$.  As discussed above, this arises from the fact that as one  decreases the mass of the 
mediator (thereby increasing the cross section as one follows
the contour lines of Fig.~\ref{Fig:DDlimitsFermion} upwards), the DM will remain in equilibrium longer. 
Eventually, for sufficiently large 
cross sections, the DM will remain in equilibrium until $T\approx m_\chi$ such 
that annihilations will begin to be significant in the non-relativistic regime. This marks the WIMP/UFO 
transition,
which is illustrated in Fig.~\ref{Fig:DDlimitsFermion} for fermionic DM, using a red shaded region 
for WIMP-like FO.

\begin{figure}[!ht]
\centering
\vskip .2in
\includegraphics[width=4.3in]{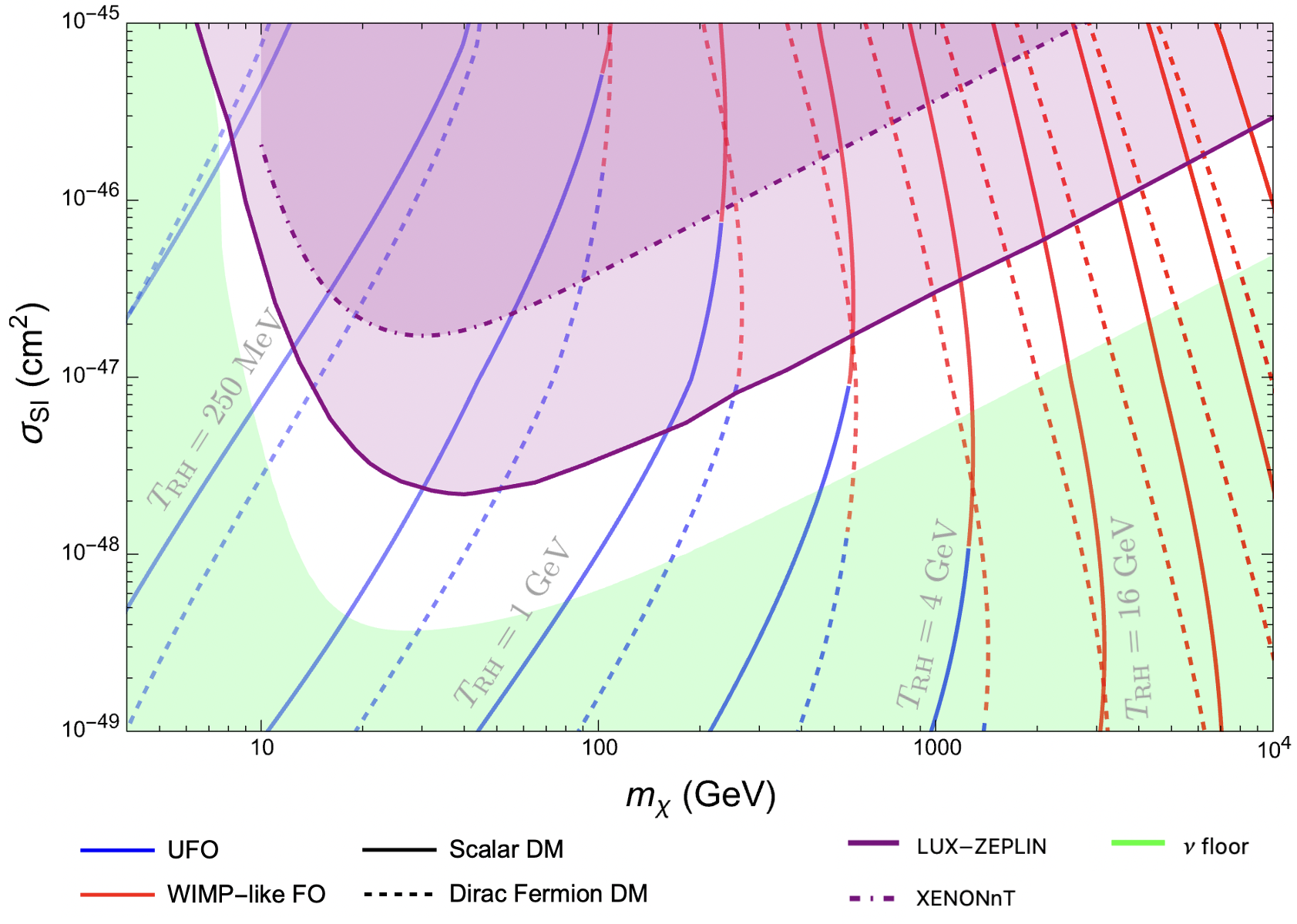}
\caption{\em \small Comparison of the available parameter space for Dirac fermion DM (dashed) vs. scalar DM (solid) produced via heavy $Z'$ vector portal interaction during reheating. Blue (red) regions of the contours correspond to UFO (non-relativistic WIMP-like FO). Direct detection limits from XENONnT \cite{XENON:2025vwd} and LUX-ZEPLIN \cite{LZ:2024zvo, LZ:2025igz} are shown in purple. The neutrino fog is shaded green.}
\label{Fig:DDlimitsFermion&Scalar}
\end{figure}

In Fig.~\ref{Fig:DDlimitsFermion&Scalar}, we compare our results for scalar DM with those of fermionic DM,
again illustrating the constraints in the ($\sigma_{\rm SI},m_\chi$) plane. The scalar DM cross sections are shown by the solid lines. The dashed lines correspond to the case of fermionic DM and are identical to the contours shown in Fig.~\ref{Fig:DDlimitsFermion}.  
The general features of the scalar curves are similar to the fermionic case. Higher UFO DM masses up to nearly 1 TeV are detectable in the scalar case, which is about 2-fold greater than the highest detectable mass for fermionic UFO DM.

For scalars, the UFO abundance is again given by Eq.~(\ref{Eq:yrhUFOmGeqTRH}), though reduced by an overall factor of 4. Then as a consequence, $\Omega_\chi h^2$ is given by Eq.~(\ref{Eq:omegaapprox}) reduced by the same factor of 4. 
As in the fermion case, we can use Eq.~(\ref{Eq:mzpsigmasi}) 
to obtain an estimate of the elastic cross section
\beq
\sigma^{\rm SI}_{\chi N} \simeq 3.7 \times 10^{-51} {\rm cm}^2 \frac{4g_{\rm RH}}{427\Sigma_{\rm tot}} \left(\frac{1~\rm GeV}{\Trh}\right)^7 \left(\frac{m_\chi}{1~\rm GeV}\right)^3 \, ,
\label{Eq:sigmaUFOs}
\eeq
 which is now a factor of 4 larger than Eq.~(\ref{Eq:sigmaUFO}).
Thus we expect the fermionic and scalar dark matter curves in the UFO region of Fig.~\ref{Fig:DDlimitsFermion&Scalar} to be parallel
with the scalar curves above the fermionic curves by a factor of 4. This relation will only strictly hold when we are deep into the UFO region, away from the UFO/WIMP transition.

The analytic analysis for WIMP-like freeze-out of scalar DM is similar to that for Dirac fermions, but differs in an important respect. As noted above, the annihilation cross section for scalar DM is p-wave suppressed. If we expand the thermally averaged cross section
as \cite{Srednicki:1988ce}
\beq
\langle \sigma v \rangle = \sum_f \frac{1}{m_\chi^2} \left(1-\frac{m_f^2}{m_\chi^2}\right)^\frac12 \left[ a_f + b_f x + \cdots \right],
\label{abexp}
\eeq

\noi
we find for scalar DM $a = 0$, and $b =  m_\chi^4/\pi M_{Z'}^4$. In this case, 
the relic density becomes 
\beq
\frac{n_\chi(a_{\rm RH})}{\Trh^3}=\frac{5 \sqrt{3}}{8}\sqrt{\frac{\pi^2 g_{\rm FO}}{30}}\frac{\Trh^3}{(45/2) b_f x_{FO}^5 m_\chi^2 M_P}\,,
\eeq
which when inserted into Eq.~(\ref{Eq:omegageneric}) gives,
\bea
\frac{\Omega_\chi h^2}{0.12} &
\simeq & 1.1 \times 10^{-11}~{\rm GeV}^{-2} \frac{\Trh^3}{ b_f m_\chi x_{\rm FO}^5}  \nonumber \\
& \simeq & 3.6 \times 10^{-11}~{\rm GeV}^{-2} \frac{M_{Z'}^4 \Trh^3}{m_\chi^5 x_{\rm FO}^5}
\label{omwimps}
\, ,
\eea
where we have again assumed all $m_f \ll m_\chi$ and again taking $g_{\rm RH}=g_{\rm FO}=427/4$. 
Once again replacing $M_{Z'}$ with the scattering cross section using Eq.~(\ref{Eq:mzpsigmasi}) and taking
$x_{\rm FO} = 1/20$, we have
\beq
\sigma^{\rm SI}_{\chi N} \simeq 1.1 \times 10^{-31} {\rm cm}^2 \left(\frac{\Trh}{1{\rm GeV}} \right)^3 \left(\frac{1{\rm GeV}}{m_\chi} \right)^5 \, ,
\eeq
leading to a cross section which is about 30 times stronger than the fermionic case. This enhancement for scalars can be seen in Fig.~\ref{Fig:DDlimitsFermion&Scalar}.

\begin{figure}[!ht]
\centering
\vskip .2in
\includegraphics[width=4.3in]{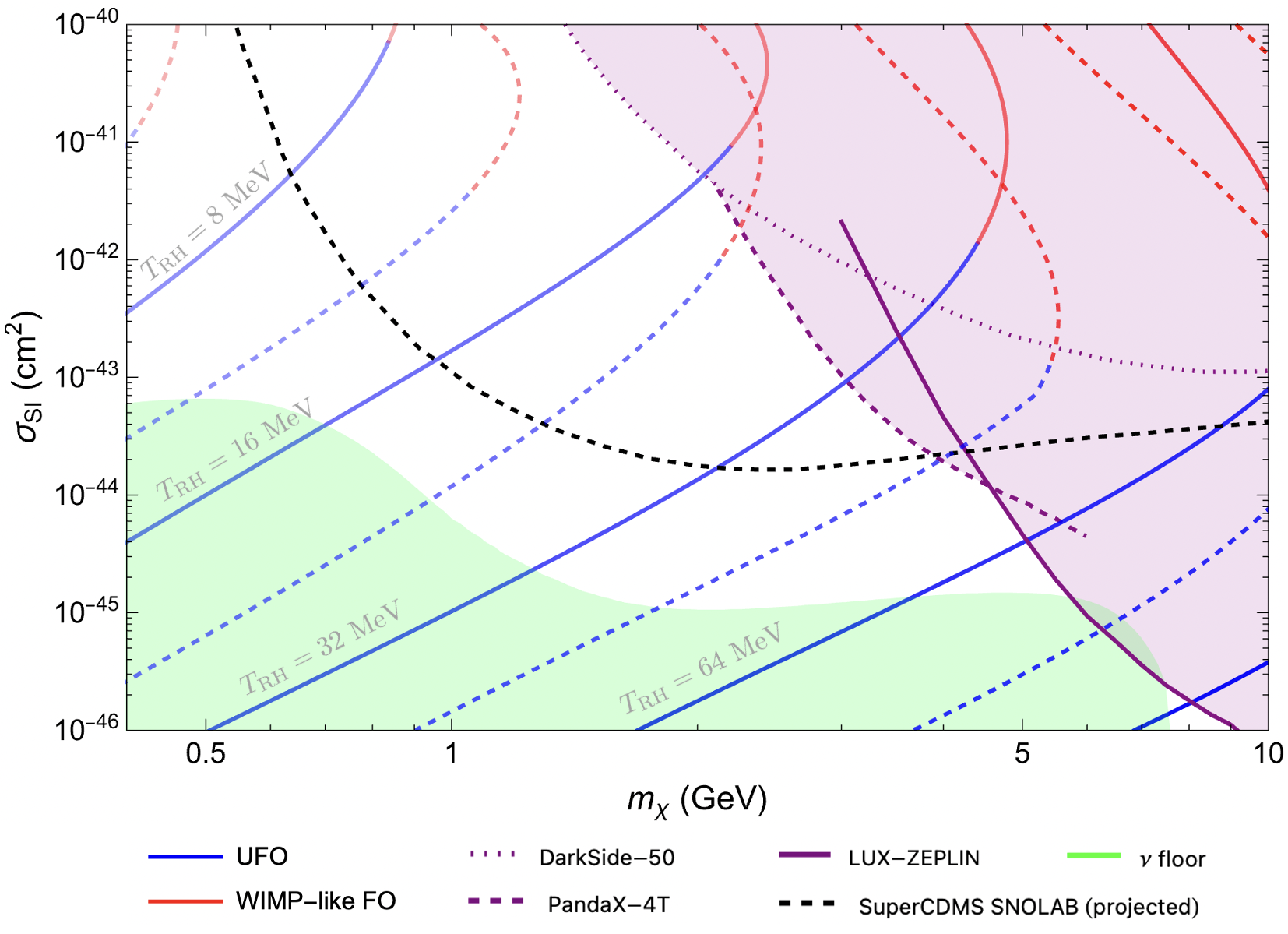}
\caption{\em \small Direct detection limits (purple) for scalar (solid) and Dirac fermion (dashed) DM in the low mass region ($0.4 \leq m_\chi \leq 10$~GeV) for DarkSide-50 \cite{DarkSide-50:2022qzh}, PandaX-4T \cite{PandaX:2025rrz}, and LZ \cite{LZ:2025igz}. The UFO and WIMP parameter space are depicted as blue and red contours respectively for fixed $T_{\rm RH}$ (the label is on the scalar contour and the fermion contour with the same temperature is below the scalar contour in the UFO regime. Projected exclusion limits for the upcoming SuperCDMS SNOLAB \cite{SuperCDMS:2022kse} experiment are also depicted (black dashed).}
\label{Fig:DDlimitsLowMass}
\end{figure}

In Fig.~\ref{Fig:DDlimitsLowMass} we show the limits from direct detection experiments for the low mass region $0.4 \leq m_\chi \leq 10$~GeV for fermionic (dashed) and scalar (solid) DM for several values of $\Trh$. The latest exclusion limits from DarkSide-50 \cite{DarkSide-50:2022qzh}, PandaX-4T \cite{PandaX:2025rrz}, and LZ \cite{LZ:2025igz} are depicted as purple shaded regions. The projected limits for SuperCDMS SNOLAB \cite{SuperCDMS:2022kse} are depicted by the black dashed curve. The contours correspond to choices of $M_{Z'}$ such that $\Omega_\chi h^2 = 0.12$ and are colored blue and red in the UFO and WIMP regimes respectively. 
In the UFO region, the fermion contours lie below the scalar contours (by a factor $\gtrsim 4$) for the temperature labeled on the scalar contour. Interestingly, the parameter space for non-relativistic freeze-out during reheating is almost entirely excluded in this region for this heavy portal model. In contrast, while a portion of the UFO parameter space is already excluded by existing limits, a large subspace remains viable, and will be probed by the forthcoming SuperCDMS SNOLAB experiment.

\subsection{Spin-dependent Scattering}

\begin{figure}[!ht]
\centering
\vskip .2in
\includegraphics[width=4.3in]{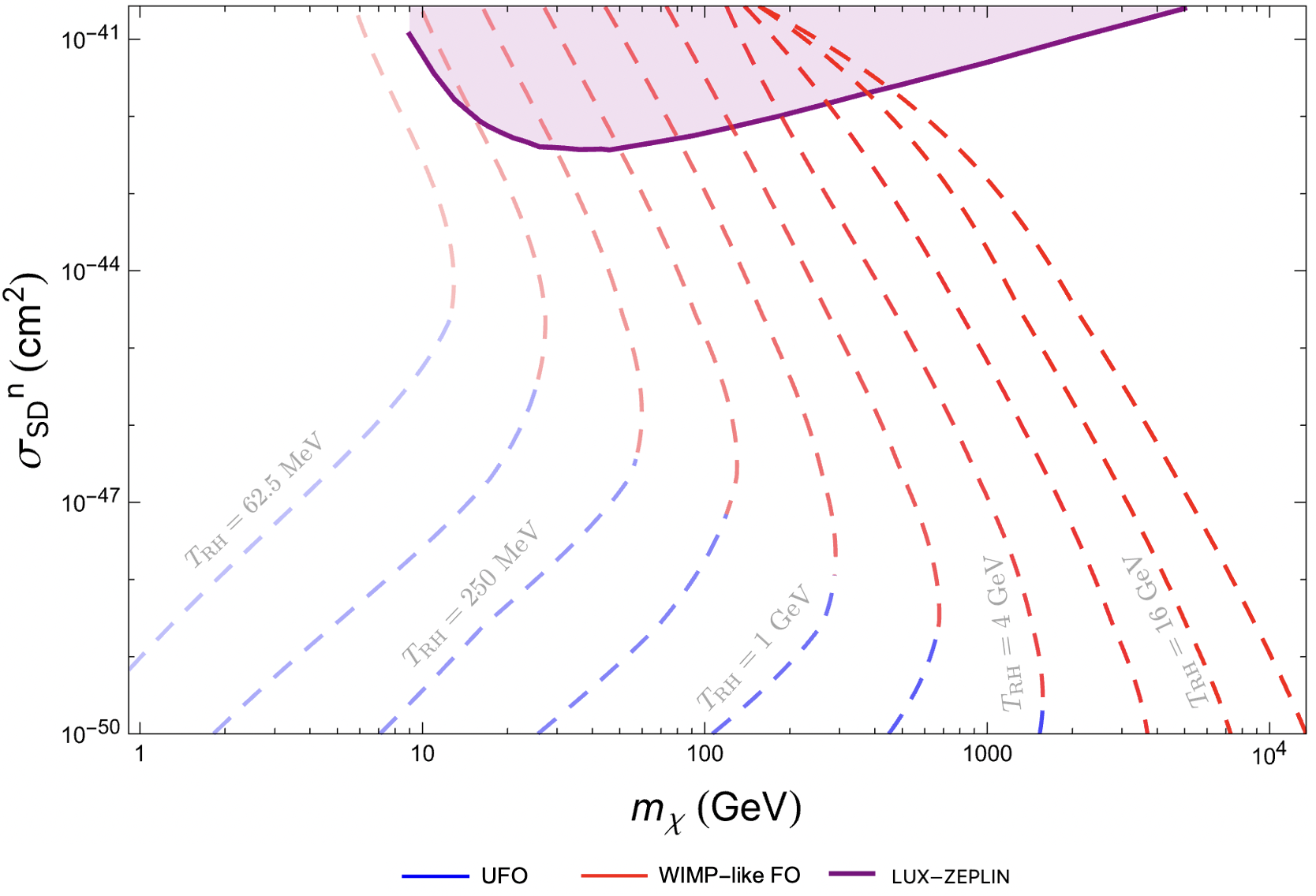}
\caption{\em \small Direct detection limits on the spin-dependent scattering cross section for scattering on neutrons from LUX-ZEPLIN (purple) for Dirac fermion UFO (blue dashed) and WIMP-like FO (red dashed) during reheating. We have used $A_q=A_\chi=1$. For each contour of fixed $T_{\rm RH}$ $M_{Z'}$ is chosen to obtain $\Omega_\chi h^2 = 0.12$ which depends on $m_\chi$.}
\label{Fig:SDnlimits}
\end{figure}

Finally we consider the case of spin-dependent scattering arising from the axial vector coupling of fermionic DM. The production of UFO DM through the axial coupling is also given by Eq.~(\ref{Eq:omegaapprox}). However, the spin-dependent DM-nucleon scattering cross section is given by Eq.~(\ref{sdsigma}) and in the limit $m_\chi \gg m_N$, 
$\mu_{\chi N} \approx m_N$ and for $A_\chi = A_q = 1$, we have
\beq
\sigma^{\rm SD}_{\chi N} \simeq \frac{3 (.29)^2 m_N^2}{\pi M_Z'^4} \simeq \frac{0.07~\rm GeV^2}{M_{Z'}^4} \,. 
\label{Eq:mzpsigmasd}
\eeq
 This is about a factor of 35 times smaller than the corresponding spin-independent cross section. Then choosing the value of $M_{Z'}$ for $\Omega_\chi h^2 = 0.12$, using Eq.~(\ref{Eq:omegaapprox}) 
\beq
\sigma^{\rm SD}_{\chi N} \simeq 2.5 \times 10^{-53} {\rm cm}^2 \frac{4g_{\rm RH}}{427\Sigma_{\rm tot}} \left(\frac{1~\rm GeV}{\Trh}\right)^7 \left(\frac{m_\chi}{1~\rm GeV}\right)^3
\label{Eq:sigmaUFOa}
\eeq
with the parametric dependence on $m_\chi$ and $\Trh$ unchanged. 
The spin-dependent cross section as a function of $m_\chi$ for 
fixed $\Trh$ is shown in Fig.~\ref{Fig:SDnlimits}.

As in the case of the vector couplings, the degeneracy in $M_{Z'}$
is also present for axial vector couplings. As a result the UFO (solid blue) curves in Fig.~\ref{Fig:SDnlimits} bend back 
to smaller masses as UFO transitions to non-relativistic freeze-out. This transition also occurs as $m_\chi$ increases. As can be seen from Eq.~(\ref{sanna}), the annihilation cross section is p-wave suppressed. While the s-wave cross section does not vanish, $a = m_\chi^2 m_f^2/2\pi M_{Z'}^4$, it is suppressed by the mass of the final state fermion. In contrast, in the limit of small final state fermion masses, $b = m_\chi^4/\pi M_{Z'}^4$. Thus, the expression (\ref{omwimps}) also holds for fermions with axial vector couplings. In that case, using Eq.~(\ref{Eq:mzpsigmasd}), we expect
\beq
\sigma^{\rm SD}_{\chi N} \simeq 3.1 \times 10^{-33} {\rm cm}^2 \left(\frac{\Trh}{1{\rm GeV}} \right)^3 \left(\frac{1{\rm GeV}}{m_\chi} \right)^5 \, ,
\eeq
in excellent agreement with the red dashed portions of the curves in Fig.~\ref{Fig:SDnlimits}. It is apparent from the figure that current experimental constraints on spin-dependent DM-nucleon scattering have not yet probed the UFO parameter space.

\subsection{Influence of $\Trh$}

\begin{figure*}[!t]
\centering
\vskip .2in
\includegraphics[width=\textwidth]{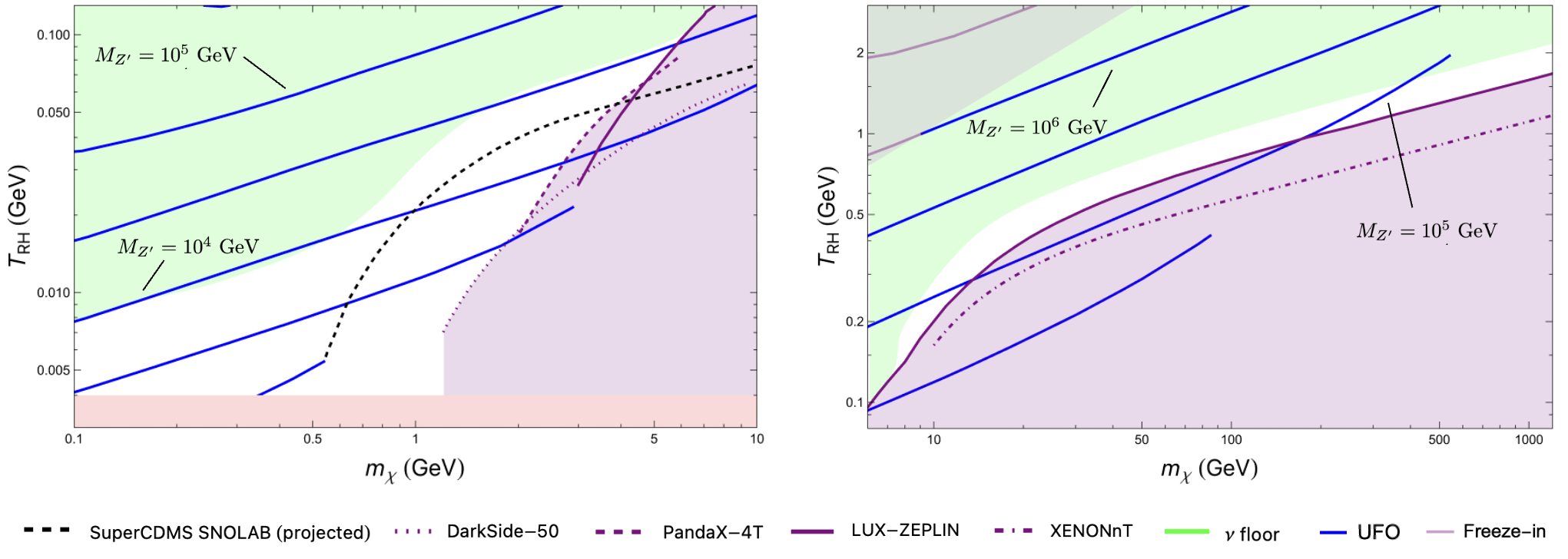}
\caption{\em \small $(m_\chi,T_{\rm RH})$ plane illustrating the viable parameter space for scalar dark matter produced through a heavy $Z'$ portal via UFO during reheating along with the regions excluded by direct detection experiments for spin-independent scattering. The white regions are viable parameter space and each blue UFO contour corresponds to a different choice of $M_{Z'}$. The green shaded region corresponds to the neutrino fog. The left (right) panel depicts the low (high) mass region. Universal vector couplings $V_\chi=V_f=1$ were used. The allowed UFO region in both plots is white while the freeze-in region (only visible in the upper left portion of the right panel) is shown as purple contour segments and shaded gray. The red shaded region in the lower left corresponds to an excluded region due to the BBN bound on $T_{\rm RH}<4$~MeV.}
\label{Fig:mTRHplane}
\end{figure*}

Finally, we can determine the constraints imposed on the  $(m_\chi, \Trh)$ plane by direct detection experiments. 
These are shown in Fig.~\ref{Fig:mTRHplane} for different mediator masses.
For high mediator masses paired with lower DM masses and higher reheating temperatures, the DM never reaches equilibrium and this freeze-in region is shaded gray, as seen in the upper left corner of the right panel of the figure.  The neutrino fog is also found in the upper left corner (of both panels) and is shaded green. The solid purple and blue contours correspond to different choices of $M_{Z'}$ for freeze-in and UFO DM candidates respectively when $\Omega_\chi h^2 = 0.12$ (the purple parts of the contours are only found in the gray-shaded region). The blue curves terminate when UFO transitions to WIMP-like FO as was seen in Figs.~\ref{Fig:UFOWIMPdegenFermionDM} and \ref{Fig:UFOWIMPdegenScalarDM} when the blue curves turn red. 
As seen in Eq.~(\ref{Eq:omegaapprox}), for a fixed value of $m_\chi$, larger 
values of $\Trh$ require a larger mediator mass to compensate for reduced dilution (between $m_\chi$ and $\Trh$ when $m_\chi > \Trh$). This ensures earlier relativistic freeze-out, and increases the time for which production occurs. This tendency
continues until the point where $M_{Z'}$ is so large that the production rate is no 
longer sufficient to compensate for the Hubble rate, thereby 
preventing dark matter from reaching thermal equilibrium and 
entering the classic FIMP regime. In the lower right corners of these figure panels, we show the current experimental constraints on the cross section as a function of the DM mass obtained from LZ (solid) \cite{LZ:2024zvo}, XENONnT (dot dashed) \cite{XENON:2025vwd}, DarkSide-50 (dotted) \cite{DarkSide-50:2022qzh}, and PandaX-4T (dashed) \cite{PandaX:2025rrz}. The white areas above these shaded regions remain to be explored in future direct detection experiments.

In the left panel of Fig.~\ref{Fig:mTRHplane}, we concentrate on the low mass region with $0.1 \le m_\chi \le 10$~GeV. Existing experimental constraints (shaded purple) have already excluded the higher DM masses in this range. The projected limits for SuperCDMS SNOLAB \cite{SuperCDMS:2022kse} depicted by the black dashed curve will further cut into the allowed region 
with relatively low reheating temperature and relatively low $Z'$ mass. It is interesting to note that the viable UFO region with $M_{Z'} < 10$~TeV may also be probed in accelerator searches. The current limits on the mass of a heavy $Z'$ from collider experiments are approximately $M_{Z'}\gtrsim 1-5$~TeV depending on the model. For a heavy $Z'$ with couplings to quarks and leptons which are identical to those of the SM Z boson, the current limit \cite{CMS:2021ctt} is $M_{Z'}>5.2$~TeV. In contrast, the current bound for a heavy $Z'$ with left right symmetric couplings ($g_L=g_R$) is $M_{Z'}>1.2$~TeV \cite{delAguila:2010mx}. The latter $Z'$ is more similar to the model investigated in this paper. In the right panel of Fig.~\ref{Fig:mTRHplane}, we concentrate on the higher mass range between 4 GeV and 1 TeV. For this mass range, direct detection experiments are closing in on the neutrino fog,currently leaving only a narrow range in $\Trh$ for a given value of $m_\chi$ for $m_{Z'} \simeq 10^5$~GeV.
As noted earlier the FIMP-like region (as well a part of the UFO region) falls within the neutrino fog area, making it particularly difficult to access for experiments such as LZ, PandaX or XENONnT.

From both panels of Fig.~\ref{Fig:mTRHplane}, we see that the region remaining to be
explored corresponds to mediators with masses of the order of $M_{Z'} \sim 1{-}300$~TeV, with reheating temperatures between 4 MeV and 2 GeV for UFO DM masses between 100 MeV and 1 TeV. 
The lower limit of $\Trh > 4$~MeV is derived from big bang nucleosynthesis \cite{tr4}. 
Reheat temperatures higher than 2 GeV would require higher values of $M_{Z'}$ pushing the cross section into the neutrino fog. 
Of course, it is not excluded that the next 
data release by teams engaged in direct detection experiments might yield a positive
signal in the coming years. 
A single ``point,'' 
corresponding to a signal in the $(m_\chi, \Trh)$ parameter space  of Fig.~\ref{Fig:DDlimitsFermion}, will translate into a reduced region in Fig.~\ref{Fig:mTRHplane}, which, in turn, will 
determine a compatible  pair $(M_{Z'}, \Trh)$ region consistent with 
the signal and the correct relic density. The direct detection 
experiment will thus become a true explorer of the primordial 
Universe, possibly providing us with an indication of the reheating temperature of the Universe.

\section{Summary}
\label{summary}

For years, many searches for dark matter have focused on a clear goal: to directly detect a candidate $\chi$ through its scattering off of electrons or nuclei.  Since the 1980s, expectations were high, as the WIMP paradigm ensured both a sufficient relic density and an achievable detection rate for dark matter masses of order 1 TeV.

However, the lack of a signal in experiments such as XENON \cite{XENON:2025vwd}, PandaX \cite{PandaX:2024qfu}, or LZ \cite{LZ:2024zvo} coupled with the lack of accelerator signatures of new particles has cast doubt on the future detectability of WIMP models. Alternatives have emerged, particularly those involving a production mechanism which does not rely on the assumption of thermal equilibrium. One of these is the FIMP \cite{fimp}. A FIMP is produced continuously from the thermal bath without ever reaching thermal equilibrium with the bath (the gravitino is a prime example \cite{grav}). Although appealing, this type of dark matter candidate relies on very weak couplings for GeV-scale dark matter, making it virtually undetectable in current scattering experiments on heavy nuclei.
The same is true for interactions proceeding via heavy vectors or scalars as in the NETDM scenario \cite{Mambrini:2013iaa,Mambrini:2015vna}. Furthermore, absent an early stage of thermal equilibrium, the issue of initial conditions remains unresolved. For example, in the case of scalar dark matter,  gravitational production may generate a significant amount of dark matter \cite{Mambrini:2021zpp,Clery:2021bwz,Cosme:2020nac,Garcia:2025rut}. The FIMP requirement for a feeble coupling can be relaxed if $m_\chi \gtrsim T_{\text{RH}}\approx T_{\rm max}$ by trading a feeble coupling for a Boltzmann suppression factor. For reheating temperatures $\lesssim 100$\,GeV such a candidate may be detectable in current direct detection experiments. This "freeze-in at stronger coupling" option \cite{Cosme:2023xpa,Arcadi:2024obp}, while resolving the detectability issue, does not solve the problem of initial conditions.

A more recent alternative which we have studied here is UFO (\textit{Ultrarelativistic Freeze-Out}) \cite{Henrich:2025gsd,Henrich:2025sli,Henrich:2025pca}. In this scenario, dark matter particles reach thermal equilibrium with the Standard Model radiation bath and decouple while relativistic, similar to case of neutrinos. While neutrinos decouple ultrarelativistically during radiation domination and would therefore typically be hot DM, UFO particles which decouple during reheating can instead be viable cold dark matter candidates. UFOs are also distinct from the standard WIMP which decouples due to Boltzmann suppression of its equilibrium number density at temperatures below its mass (non-relativistic freeze-out). UFO decoupling instead occurs because the DM-SM interactions becomes too weak to sustain chemical equilibrium during the reheating period after inflation. In this case, there may be enough time between decoupling and $T_{\text{RH}}$ to "drown" dark matter in a thermal bath, thus respecting $n_\chi/n_\gamma \simeq 10^{-9}(\frac{1 \text{ GeV}}{m_\chi})$. This model avoids the problem of initial conditions thanks to the existence of thermal equilibrium, while maintaining a sufficiently strong coupling with the visible sector to be potentially observable today.

In this paper, we obtained and analyzed constraints on fermionic and scalar UFO dark matter arising from the non-observation of signals in direct detection experiments. We focused on a UV model where the mediator between the visible sector and dark matter is a heavy $Z'$ boson. Our main results are illustrated in Figs.~\ref{Fig:DDlimitsFermion&Scalar} and \ref{Fig:DDlimitsLowMass}, where we plotted iso-$T_{\text{RH}}$ curves, clearly distinguishing the UFO regime (in blue) from the WIMP regime (in red). We found that ongoing direct detection experiments have already excluded large regions of UFO parameter space. In the high mass region, UFO DM candidates with masses $8~{\rm GeV}\lesssim m_\chi \lesssim 800$~GeV remain viable above the neutrino fog for low reheating temperatures $125 \text{ MeV} \lesssim T_{\rm RH} \lesssim 3$~GeV. For light dark matter (below 10 GeV), we found that direct detection experiments have nearly excluded DM produced by non-relativistic (WIMP-like) freeze-out in our heavy portal model. In contrast, there remains a large region of viable UFO parameter space for light DM masses $0.4~{\rm GeV} \lesssim m_\chi \lesssim 6$~GeV, much of which will be accessed by the upcoming SuperCDMS SNOLAB experiment. While the UFO mechanism is generally compatible with a wide range of heavy mediator masses $10^3 \text{ GeV} \lesssim M_{Z'} \lesssim 10^{14}$~GeV \cite{Henrich:2025gsd,Henrich:2025sli}, we found that values of $M_{Z'}$ on the lower end of this spectrum ($1~{\rm TeV} \lesssim M_{Z'}\lesssim 300$~TeV) are compatible with dark matter that is accessible to direct detection experiments. Moreover, collider searches will provide a complementary approach to ruling out UFO DM candidates produced by heavy mediators with $1~{\rm TeV} \lesssim M_{Z'} \lesssim 10$~TeV.

In conclusion, our study opens a window into the use of direct detection experiments as cosmic explorers. While WIMP models are under significant tension, and FIMP models face issues related to initial conditions and/or challenging detectability, the UFO scenario appears to be one of the most promising (and natural) alternatives. Many alternative scenarios will, in any case, require a deeper analysis of dark matter-visible sector interactions during the pre-BBN era including reheating, thereby transforming the quest for a dark matter particle into constraints on the physics of the early Universe.

\acknowledgments
The authors would like to thank Mathieu Gross, Olivier Deligny and Xavier Bertou for helpful discussions.
This project has received support from the European Union's Horizon 2020 research and innovation program under the Marie Sklodowska-Curie grant agreement No 860881-HIDDeN.
  The work of K.A.O.~was supported in part by DOE grant DE-SC0011842  at the University of
Minnesota. Y.M. acknowledges support by Institut Pascal at Université Paris-Saclay during the Paris-Saclay Astroparticle Symposium 2025, with the support of the P2IO Laboratory of Excellence (program “Investissements d’avenir” ANR-11-IDEX-0003-01 Paris-Saclay and ANR- 10-LABX-0038), the P2I axis of the Graduate School of Physics of Université Paris-Saclay, as well as the CNRS IRP UCMN.

\end{document}